\documentclass{article}

\usepackage{arxiv}
\usepackage[utf8]{inputenc} %
\usepackage[T1]{fontenc}    %
\usepackage{hyperref}       %
\usepackage{url}            %
\usepackage{booktabs}       %
\usepackage{amsfonts}       %
\usepackage{nicefrac}       %
\usepackage{microtype}      %
\usepackage{graphicx}
\usepackage{natbib}
\usepackage{doi}            %

\usepackage{algorithm}
\usepackage{algpseudocode}
\usepackage{listings}
\usepackage{xcolor}
\usepackage{mathtools}
\usepackage{amsmath}        %
\usepackage{amsthm}         %
\usepackage[capitalize,nameinlink]{cleveref}

\usepackage{booktabs}       %
\usepackage{algorithm}      %
\usepackage{algpseudocode}  %
\usepackage{listings}       %
\usepackage{xcolor}
\usepackage{mathtools}
\usepackage[utf8]{inputenc}
\usepackage{threeparttable}
\usepackage{tikz}
\usetikzlibrary{positioning, arrows.meta, backgrounds, calc, shapes.geometric, shadows.blur}
\usepackage{hyperref}
\usepackage[capitalize,nameinlink]{cleveref}
\usepackage{xurl}

\theoremstyle{plain}
\newtheorem{theorem}{Theorem}[section]
\newtheorem{lemma}[theorem]{Lemma}
\newtheorem{proposition}[theorem]{Proposition}

\theoremstyle{definition}

\title{Sell Data to AI Algorithms Without Revealing It: Secure Data Valuation and Sharing via Homomorphic Encryption}

\date{} 					%

\author{
  Michael Yang \\
  University of Texas at Dallas\\
  \texttt{YunxuanMichael.Yang@UTDallas.edu} \\
  \And
  Ruijiang Gao \\
  University of Texas at Dallas\\
  \texttt{Ruijiang.Gao@UTDallas.edu} \\
  \And
  Zhiqiang (Eric) Zheng \\
  University of Texas at Dallas\\
  \texttt{ericz@utdallas.edu} \\
}

\begin{document}
\maketitle

\begin{abstract}
The rapid expansion of Artificial Intelligence is hindered by a fundamental friction in data markets: the ``value-privacy dilemma'', where data buyers cannot verify data utility without inspection, yet inspection may expose the data. We resolve this challenge by introducing the Trustworthy Influence Protocol (TIP), a privacy-preserving framework that enables prospective buyers to quantify the utility of external data without ever seeing the raw assets. By integrating Homomorphic Encryption with gradient-based influence functions, our approach allows for a precise, blind scoring of data  against a buyer's specific AI model. To ensure scalability for foundational models such as Large Language Models (LLMs), we employ low-rank gradient projections that reduce computational overhead while maintaining near-perfect fidelity to plaintext baselines, as demonstrated across BERT and GPT-2 architectures. Empirical simulations show that encrypted valuation signals achieve a high correlation with the ground-truth data utility and reveal a concentrated distribution of data value in pre-training corpora where only a small portion of data typically helps improve the model performance. These findings challenge the prevailing flat-rate data pricing mechanism used in legal practice and offer a scalable technical foundation for a meritocratic, secure data economy. A complete implementation of our algorithms for replication is available at an anonymous dedicated GitHub repository: \url{https://anonymous.4open.science/r/FHELLM-C1E1/README.md}.
\end{abstract}


\section{Introduction}

Advanced machine learning models require large, high-quality datasets to learn effectively \citep{zhang2020survey}. For example, Llama-3 was pretrained on over 15 trillion tokens of data from publicly available sources like Wikipedia and Reddit~\citep{meta2024llama3}. Recognizing that generic web-scale pretraining data often fails to capture domain-specific knowledge, AI models increasingly seek proprietary datasets to gain specialized capabilities. In parallel, AI companies now offer standardized fine-tuning pipelines, making it operationally straightforward to incorporate new proprietary data into deployed models \citep{openai2024finetuning}.

Historically, most AI foundation models started with mass scraping publicly accessible content without authorization, triggering a wave of copyright disputes. 
In the United States, Getty Images sued Stability AI, alleging that millions of copyrighted photographs were incorporated into its generative model's training corpus without a license \citep{Brittain2023_Getty_Reuters}. In \emph{Bartz vs.\ Anthropic} lawsuit, authors represented by Andrea Bartz challenged Anthropic's use of copyrighted books, and the dispute moved toward a class-wide settlement awarding rights holders a largely uniform payment of roughly \$3,000 per covered work: a salient example of ``flat-rate'' compensation \citep{AnthropicSettlement_FAQ,AnthropicSettlement_Dates}. In February 2025, a U.S. federal court ruled that Ross Intelligence infringed Westlaw's copyrighted headnotes and rejected its fair-use defense \citep{Bibas2025_ThomsonReuters_v_Ross_MemorandumOpinion}. Judicial scrutiny has intensified globally: in Germany, OpenAI was sued for unauthorized memorization and reproduction of protected song lyrics, in violation of EU copyright law \citep{LGMunichI2025_GEMA_OpenAI_Memorisation_Judgment}; in China, the Hangzhou Internet Court held an unnamed generative AI platform liable for training on protected anime character models and ordered it to compensate the plaintiff for 30,000 RMB in damages \citep{HangzhouInternetCourt2024_Ultraman_GenAIPlatform}.
These cases represent some of the earliest judicial examinations of whether commercial AI training on proprietary data constitutes copyright infringement.

Facing mounting legal pressure, many AI companies have begun securing formal data-licensing agreements. Reddit established licensing partnerships with AI companies such as OpenAI and Google \citep{OpenAI2024_RedditPartnership}; Meta struck commercial deals with major news publishers \citep{reuters2025meta}; OpenAI licensed part of the Associated Press archive \citep{AP2023_OpenAI_APArchive} and formed multi-year partnerships with Axel Springer, TIME, and News Corp \citep{Reuters2023_OpenAI_AxelSpringer,OpenAI2024_TIME_Partnership,NewsCorp2024_OpenAI_Partnership}. Yet unauthorized use persists even as the law adapts. 
In February 2026, ByteDance launched Seedance 2.0, an AI video-generation model that immediately drew cease-and-desist letters from Disney, Paramount, Warner Bros., Netflix, and Sony for reproducing their copyrighted characters without authorization \citep{reuters_bytedance_ai_video_2026,thewrap_disney_bytedance_seedance_2026}.
Similarly, the Motion Picture Association sued a major generative AI company, characterizing the infringement as ``systemic'' that baked into the model itself \citep{mpa_seedance_cease_2026}. These incidents illustrate that the incentive to train on unlicensed data remains unfazed.

A complication to these unsettled phenomena arises from a fundamental economic friction: even when AI companies do seek to license data legitimately, the value of information cannot be verified by the data buyer without first receiving it, yet receiving it already transfers the data asset before paying. This friction is known as Arrow's Information Paradox \citep{arrow1962economic}, one that formal agreements alone cannot resolve. In the context of selling data to AI algorithms, a buyer cannot assess how much a dataset will improve their model without inspecting it or training on it, but doing so inherently exposes the data and creates a significant risk for the data owner. Conversely, the buyer risks adverse selection, potentially overpaying for data whose quality is exaggerated or manipulated by the seller \citep{zhang2024survey}.

One potential workaround is to introduce a trusted broker to facilitate the process. For instance, a neutral escrow service or data trust organization could hold the sensitive dataset, run the valuation on behalf of the buyer, and report the results. In practice, however, this approach shifts the trust problem rather than solving it. \citet{spiekermann2019data} note that a ``lack of trust and security'' in intermediaries is a major obstacle to data sharing. The data owner must trust the broker not to leak or misuse their intellectual property, while the buyer must trust that the evaluation is accurate and unbiased.

\begin{figure}[htbp]
    \makebox[\textwidth][c]{%
        \resizebox{\textwidth}{!}{%
            \begin{tikzpicture}[
                    box/.style={draw, rounded corners=6pt, text width=7cm, minimum height=1.8cm, align=center, very thick, font=\huge\sffamily, fill opacity=0.95, blur shadow={shadow blur steps=5, shadow xshift=3pt, shadow yshift=-3pt, shadow opacity=20}},
                    buyer/.style={box, top color=white, bottom color=blue!8, draw=blue!80!black},
                    seller/.style={box, top color=white, bottom color=green!8, draw=green!70!black},
                    broker/.style={box, top color=white, bottom color=orange!8, draw=orange!90!black},
                    legend_box/.style={box, top color=white, bottom color=gray!5, draw=gray!60, text width=19cm, font=\huge\sffamily, align=left},
                    line/.style={
                        -{Latex[length=4mm, width=3mm]}, 
                        line width=1.2pt,                
                        draw=black!75,                   
                        rounded corners=6pt,             
                        preaction={
                            draw=black, opacity=0.15, line width=3pt, transform canvas={xshift=1.5pt, yshift=-1.5pt} 
                        }
                    },
                    dashed_arrow/.style={-{Latex[length=4mm, width=3mm]}, thick, dashed, draw=black!60, rounded corners=6pt},
                    dashline/.style={-{Latex[length=4mm, width=3mm]}, line width=1.2pt, dashed, draw=blue!60, rounded corners=6pt},
                    lane_label/.style={font=\huge\sffamily\bfseries, align=center}
                ]

                \node[lane_label, text=green!60!black]  at (-6, 6)  {\includegraphics[width=2.5cm]{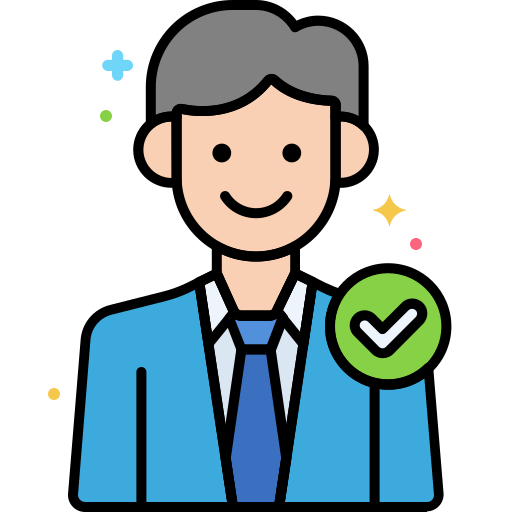}\\[3mm]Data Seller};
                \node[lane_label, text=orange!80!black] at (-6, 0)  {\includegraphics[width=2.5cm]{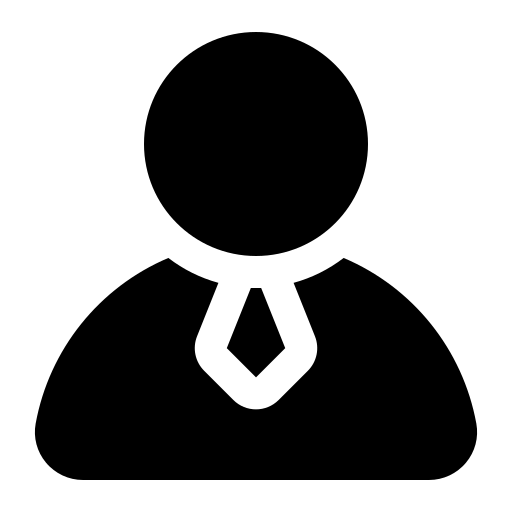}\\[3mm]Broker};
                \node[lane_label, text=blue!70!black]   at (-6, -6) {\includegraphics[width=2.5cm]{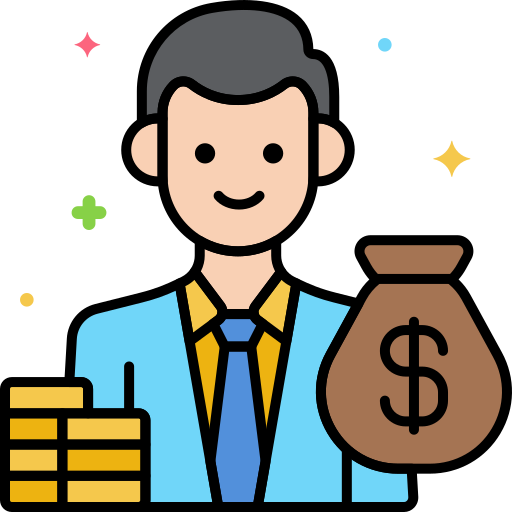}\\[3mm]Data Buyer};
            
                \draw[thick, black!60, dashed] (-8, 3.3) -- (39, 3.3);
                \draw[thick, black!60, dashed] (-8, -3.3) -- (39, -3.3);

                \node[buyer] (b0) at (1, -6) {\textbf{Phase 1 (i): Key Setup}\\ 
                    \includegraphics[width=8mm]{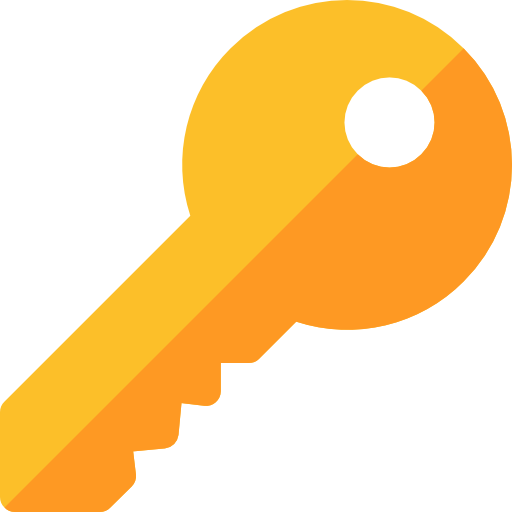} $pk$, 
                    \includegraphics[width=8mm]{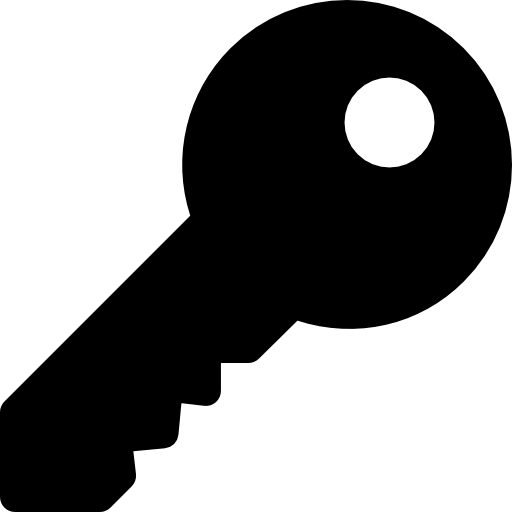} $sk$, 
                    \includegraphics[width=8mm]{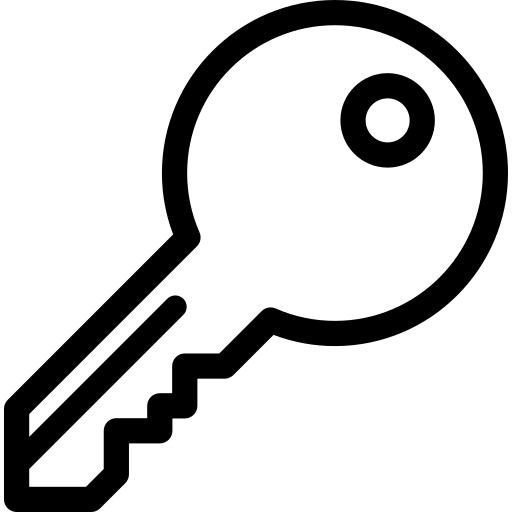} $evk$};
            
                \node[broker, text width=5cm] (br0) at (8.5, 0) {\textbf{Store}\\ \includegraphics[width=8mm]{evaluationkey.png} $evk$}; 
            
                \node[seller] (s1) at (9.5, 6)  {\textbf{Phase 2 (i): Inference \& Gradient}\\ 
                    $\nabla_\theta \ell(z_s; \hat{\theta})$};
                    
                \node[buyer]  (b1) at (9.5, -6) {\textbf{Phase 1 (ii): Inference \& Preconditioning}\\ 
                    $\nabla_\theta \ell(z_{eval})^\top H_{\hat{\theta}}^{-1}$};
            
                \node[seller, text width=9cm] (s2) at (18.5, 12.5)  {\textbf{Phase 2 (ii): Encrypt Gradient}\\ 
                    $\text{Enc}_{\mathsf{pk}}\big(\nabla_\theta \ell(z_s; \hat{\theta})\big)$\\ \vspace{1mm}\huge\texttt{b'\textasciicircum{}\textbackslash x10\textbackslash xa3...}};
                    
                \node[buyer, text width=9cm]  (b2) at (18.5, -12.5) {\textbf{Phase 1 (iii): Encrypt Vector}\\ 
                    $\text{Enc}_{\mathsf{pk}}\big(\nabla_\theta \ell(z_{eval})^\top H_{\hat{\theta}}^{-1}\big)$\\ \vspace{1mm}\huge\texttt{b'\textasciicircum{}\textbackslash xa1\textbackslash x09...}};
            
                \node[broker, text width=9cm] (br3) at (24, 0) {\textbf{Phase 3: FHE Influence Calculation}\\ 
                    $\text{Enc}_{\mathsf{pk}}\big(\nabla_\theta \ell(z_{eval})^\top H_{\hat{\theta}}^{-1}\big)$\\ $\otimes$\\ 
                    $\text{Enc}_{\mathsf{pk}}\big(\nabla_\theta \ell(z_s; \hat{\theta})\big)$};
            
                \node[broker, text width=7cm] (br4) at (34.5, 0) {\textbf{Return Encrypted Score}\\ 
                $\mathsf{ct}_{\mathrm{score}}$};
            
                \node[buyer, text width=7cm] (b5) at (34.5, -6.3) {\textbf{Phase 4: Decrypt and Score}\\ $\hat{s}(z_s)=\text{Dec}_{\mathsf{sk}}(\mathsf{ct}_{\mathrm{score}})$};
            
                \node[legend_box, text width=13cm] (legend) at (32, 9) {
                    \textbf{Cryptographic Keys:} \\ 
                    \includegraphics[width=7mm]{publickey.png} \textbf{$pk$}: Public Key (for Encryption) \\ 
                    \includegraphics[width=7mm]{privatekey.png} \textbf{$sk$}: Private Key (for Decryption) \\ 
                    \includegraphics[width=7mm]{evaluationkey.png} \textbf{$evk$}: Evaluation Key (for Relinearization)
                };

                \draw[line] (b0.north) |- node[above, pos = 0.7, font=\huge\sffamily] {Share \includegraphics[width=10mm]{publickey.png} $pk$} (s1.west);
                \draw[line] (b0.north) |- node[right, pos=0.25, font=\huge\sffamily] {Forward \includegraphics[width=10mm]{evaluationkey.png} $evk$} (br0.west);
            
                \draw[line] (b0.east) -- (b1.west);
                
                \draw[line] (s1.north) |- node[left, font=\huge\sffamily] {Encrypt with \includegraphics[width=10mm]{publickey.png} $pk$} (s2.west);
                \draw[line] (b1.south) |- node[left, font=\huge\sffamily] {Encrypt with \includegraphics[width=10mm]{publickey.png} $pk$} (b2.west);
            
                \draw[line] (s2.east) -| node[left, pos=0.8, font=\huge\sffamily] {Forward to Broker} (br3.north);
                \draw[line] (b2.east) -| node[left, pos=0.8,  font=\huge\sffamily] {Forward to Broker} (br3.south);
            
                \draw[dashed_arrow] (br0.east) -- node[above, font=\huge\sffamily, text=orange!80!black] {For Relinearization} (br3.west);
            
                \draw[line] (br3.east) -- (br4.west);
                \draw[line] (br4.south) -- (b5.north);
            
                \draw[dashline] (b0.south) -- ++(0, -1.5) -| node[above, near start, font=\huge\sffamily, text=blue!80!black] {Keep \includegraphics[width=6mm]{privatekey.png} $sk$ for decryption} (b5.south);
            
                \end{tikzpicture}%
        }%
    }
    \caption{Workflow of the Trustworthy Influence Protocol (TIP) for the secure data marketplace. The Buyer's three sub-steps (Phase 1) and the Seller's two sub-steps (Phase 2) proceed in parallel; the Broker then performs the homomorphic evaluation (Phase 3); and the Buyer decrypts the final scores (Phase 4), as formalized in Section~\ref{sec:he_ckks}.}
    \label{fig:fhe}
\end{figure}

In light of these challenges, we propose an encrypted data-valuation framework that integrates privacy-preserving computation via Homomorphic Encryption (HE) with data valuation methods. 
HE is a cryptographic approach that allows computation to be performed directly on encrypted data (ciphertexts), such that decrypting the output yields the same result (or a close approximation) as computing on the underlying unencrypted data \citep{marcolla2022survey}. 
In our framework, the buyer first produces an encrypted scoring task by HE that represents what the model needs to improve, computed from the buyer’s private evaluation data, while the seller converts each candidate data point into a compact, encrypted representation. An untrusted broker then performs the scoring computation directly on these encrypted inputs and returns only encrypted scores. 
Finally, only the buyer decrypts the returned scores to learn how useful each candidate data point would be, while the seller and the broker never sees the buyer’s private evaluation set, while the buyer and the broker never sees the seller's raw data.
The score is based on a gradient influence approximation that quantifies the marginal change in the buyer's evaluation loss if the candidate data point were incorporated into training. The full protocol and its cryptographic components are illustrated in \Cref{fig:fhe} and formalized in \Cref{sec:method}. As detailed in \Cref{sec:exp}, this framework recovers the utility of data with near-perfect accuracy and minor computing overhead, a property especially critical for large foundational models such as Large Language Models (LLMs).

Our work makes the following contributions. 

\begin{itemize}
    \item First, we design a novel encrypted data-valuation framework that integrates gradient-based influence functions with Homomorphic Encryption. We introduce the Trustworthy Influence Protocol (TIP), which enables buyers to compute precise utility scores on encrypted gradients without ever exposing their model parameters or viewing the raw data. 
    \item Second, to demonstrate generality, we evaluate TIP across three representative model families that span over increasing scale and complexity: a Multilayer Perceptron (MLP) for image data, Bidirectional Encoder Representations from Transformers (BERT) for text classification, and a Generative Pre-trained Transformer (GPT) for text generation. 
    In all cases, encrypted scoring produces essentially the same values and the same purchase ranking as the standard computation one would obtain if the data were fully visible in a trusted environment, while incurring only a marginal computational overhead, making the approach practical even for large models.
    \item Third, we further validate the approximation consistency of our valuation signal through a healthcare data market built with real data. By comparing encrypted scores against ground-truth model retraining in a regulated inpatient setting, we show that our model-based metric achieves a 0.96 correlation with realized utility, significantly outperforming standard similarity-based heuristics and effectively resolving adverse selection risks in high-stakes environments. 
    \item Finally, we apply our framework to a Generative AI book market to uncover the economic implications of data heterogeneity. The numerical experiments reveal that data utility is highly skewed, with a minority of texts driving capability while the majority degrades performance (see \Cref{fig:bookmkt}), demonstrating that the existing ``flat-rate'' compensation methods fail to capture the true marginal value of data, and establishing the necessity for a meritocratic, privacy-preserving data marketplace.
\end{itemize}

The remainder of the paper is organized as follows. \Cref{sec:related_work} discusses the relevant literature in data markets, data attribution, and privacy-aware computing.
\Cref{sec:method} formalizes our solution. We first define the market environment and the economic constraints of the AI data marketplace. We then introduce the mathematical foundations of influence-function-based valuation and construct TIP, detailing how HE resolves the value-privacy dilemma. We conclude this section with formal proofs of the protocol's security properties.
In \Cref{sec:exp}, we empirically validate the framework. We demonstrate its approximation consistency and computational scalability across diverse model architectures, from MLP to GPT-2. We then instantiate the protocol in two real-world applications: a healthcare data exchange and a generative AI book market to prove its approximation consistency and illustrate the necessity of transitioning from flat-rate pricing to meritocratic compensation.
\Cref{sec:discussion} details the boundary conditions of our approach, including constraints on the acquisition size and model accessibility. We then outline pathways for practical deployment.
Finally, \Cref{sec:conclusion} concludes the paper with a summary of our contributions to the design of efficient and secure data economies. All proofs and supplemental materials are presented in the appendices.

\section{Related Work}
\label{sec:related_work}

In this section, we position our work within three strands of literature. We first review research on data markets and data exchanges, emphasizing the persistent value–privacy dilemma that undermines trust between data buyers and sellers. We then survey feature attribution methods and data attribution methods, including Shapley-based approaches and influence-function-based techniques, and explain why the latter are particularly suitable for model-specific data valuation at scale. Finally, we discuss privacy-aware computation, contrasting differential privacy and secure multi-party computation with Homomorphic Encryption (HE), and explain how our use of HE fills a gap between scalable data valuation and strong cryptographic guarantees.

\subsection{Data Markets}

The emergence of data marketplaces, which facilitate the exchange of data between sellers and buyers, has prompted a substantial body of research on their design, pricing mechanisms, and trust infrastructure \citep{zhang2024survey}. Unlike traditional commodities, data exhibits a set of unique economic properties: it is non-rivalrous, easily replicable, highly context-dependent in value, and difficult to evaluate ex ante. These features render standard market models inadequate and necessitate tailored mechanisms for pricing and trust.

Early research focused on structural paradigms of data exchange, including direct data sales, query-based access, and trained-model transactions. For instance, \citet{balazinska2011data} and \citet{koutris2015query} proposed systems where buyers issue queries over sellers' data without needing full access to raw content, while  \citet{agarwal2019marketplace} introduced a formal auction-based market for trading machine learning models derived from different training data. These studies emphasized the need for platforms to mediate between data contributors and model consumers, yet assumed either transparent trust between parties or a third party to enforce contracts.

A central challenge cutting across these designs is the well-known value-privacy dilemma: the buyer cannot assess the true utility of data without interacting with it, but such interaction inherently reveals the data's content, risking theft or misuse. This dilemma is a practical manifestation of Arrow’s Information Paradox \citep{arrow1962economic}, long recognized in information economics. Recent works have addressed this problem from two angles. The first involves economic mechanisms for fair pricing under uncertainty. \citet{roughgarden2010algorithmic} and \citet{wang2021blockchain} investigate combinatorial pricing and blockchain-based incentive systems to ensure fair compensation and traceability. \citet{spiekermann2019data} and \citet{fernandez2020data} studied real-world platform dynamics, showing that the lack of enforceable privacy guarantees deters agents from joining data marketplaces despite their potential economic benefits.

A second line of work embeds privacy constraints directly into the market design. \citet{ghosh2011selling} introduced the concept of pricing private data via differential privacy, proposing payment schemes that compensate individuals based on the privacy loss incurred. \citet{li2014theory} and \citet{chen2019towards} developed arbitrage-free pricing models where buyers purchase noisy query answers or differentially-private models rather than raw data. These frameworks conceptualize privacy-aware pricing, but do not provide accurate utility measurement at the data level nor scale to high-dimensional learning tasks.

We bridge the gap between the economic necessity of verifying utility and the technical requirement of confidentiality. 
In our data market design, valuation precedes purchase, allowing buyers to estimate the exact marginal utility of a candidate data while preserving both parties’ privacy. Unlike model-free metrics designed for feature selection in federated alliances, our influence-based approach is model-specific, directly quantifying the reduction in loss for the buyer's target task—a metric more aligned with the ``willingness to pay'' in open markets. Furthermore, our HE framework offers a software-based cryptographic root of trust, allowing buyers to compute precise utility scores on encrypted data without ever exposing the raw assets.

\subsection{Data Attribution Methods}

Data attribution methods quantify the model's behavior change if specific training instances were removed, modified, or re-weighted \citep{jia2019towards}.
Data attribution evaluates the utility of individual training samples, fundamentally differing from feature attribution methods like LIME or SHAP \citep{NIPS2017_7062,ribeiro2016should}. 
The stakes of this inquiry have risen sharply with the deployment of Generative AI. For example, OpenAI was accused for deleting its ChatGPT training data to disguise the value of publisher (including NYT) data after being sued for copyright violations \citep{Salman2024_OpenAI_Deleted_TrainingData}. Attribution now serves as the technical substrate for resolving copyright disputes, enabling machine unlearning of toxic content \citep{yao2024machine}, and calculating fair compensation for content creators \citep{grynbaum2023times}.

Data attribution methods generally fall into two categories: retraining-based and gradient-based \citep{deng2025survey}. Foundational approaches like the Data Shapley Value \citep{ghorbani2019data} provide a theoretically rigorous framework for fair valuation but are computationally prohibitive for large models, as they require retraining the model on a combinatorial set of data samples. To circumvent retraining, Influence Functions (IF) utilizes a first-order Taylor expansion to approximate the effect of data on the model parameters \citep{koh2017understanding}. Formally, the influence of a training point $z$ on a test point $z_{test}$ is computed via the Inverse Hessian-Vector Product (IHVP). 
This geometric approach captures not just the directional alignment of gradients, but also the curvature of the loss landscape, effectively down-weighting redundant data in curved (high-curvature) directions and rewarding novel data in flat (low-curvature) directions.

However, calculating the inverse Hessian for billion-parameter models is intractable. Recent research has focused on scalable approximations. For example, using second-order optimization techniques can efficiently approximate the Hessian \citep{pearlmutter1994fast, martens2010deep, agarwal2017second}. 
TRAK \citep{park2023trak} replaces the Hessian with a linearized gradient similarity metric using random projections. Low-rank Gradient projection (LoGra, \citep{choe2024your}) further exploits the low-rank structure of gradients in linear layers to reduce dimensionality. Most recently, \citet{hu2025grass} achieved sub-linear time complexity by leveraging the inherent sparsity of gradients combined with sparse Johnson-Lindenstrauss transforms (SJLT), outperforming previous baselines in throughput.

Our work differentiates itself from this literature in three fundamental ways. First, the application of influence functions in data attribution is retrospective: they analyze existing training data to answer, ``Which samples were responsible for this prediction?'' In contrast, our framework is prospective: we treat the seller's data as a candidate $z_{new}$ that does not yet exist in the buyer's model. We adapt the influence formalism to estimate the counterfactual: ``How much would the evaluation loss decrease if this unseen data were added to the training set?'' This effectively repurposes attribution from a debugging tool into a marginal utility function for data acquisition.
Second, the influence-function formula possesses a structural property that makes it uniquely amenable to privacy-preserving computation: it naturally decomposes into two independent components, one derived entirely from the buyer's private assets and the other derived entirely from the seller's candidate data. The final evaluation of the data only depends on the inner product of these two components. This clean separation allows each party to prepare its own component in isolation, and an untrusted intermediary to combine them under encryption without ever observing either input. Alternative valuation methods lack this decomposability. 
Third, while standard attribution methods assume full transparency of both the model parameters and the training corpus, our framework adapts these metrics to operate entirely within a cryptographic environment. By bridging Homomorphic Encryption with gradient-based influence, we resolve the trust deficit inherent in open data exchanges.

\subsection{Information Privacy and Privacy-Aware Computing}

Our work contributes to the Information Systems literature on information privacy \citep{SmithDinevXu2011, XuDinev2022}. Within this domain, a behavioral stream has established that privacy concerns are context-dependent and shaped by perceived control and assurance mechanisms \citep{XuTeoTanAgarwal2012, BuckmanBockstedtHashim2019, XuZhang2022}, while a technical stream develops computational methods that enable data sharing without exposing sensitive content \citep{LiSarkar2011, LiQin2017, HanWangWuFang2025}. For instance, one line of technical approaches that focus on anonymizing or sanitizing data for release \citep{menon2016privacy, ghoshal2020hiding}; a related strand addresses privacy risks on the model-sharing side, where adversaries may infer confidential properties of a provider's training data from shared model parameters or outputs \citep{yang2025secure}. 
Our study extends this technical tradition by introducing a Homomorphic Encryption framework that resolves a privacy challenge specific to AI data markets: enabling buyers to verify the utility of data for their models without the data ever leaving encrypted form. 
Our framework preserves the full analytical signal needed for model-specific valuation while providing cryptographic confidentiality guarantees.

Resolving the value-privacy dilemma requires a computation method that allows the buyer to verify data utility without ever viewing the raw asset. The dominant paradigm in privacy-preserving machine learning is Differential Privacy \citep{dwork2014algorithmic}, which protects individual data by injecting calibrated noise into the computation outputs or gradients. In the context of data valuation, \citet{wang2023security} proposed TKNN-Shapley, a privacy-friendly valuation metric designed to bound sensitivity and accommodate DP noise. However, DP presents a fundamental conflict for data markets: the ``privacy-utility trade-off''. In a commercial transaction, the buyer needs a precise signal of marginal utility to determine willingness to pay. The noise required to satisfy rigorous DP guarantees directly degrades the precision of this valuation score \citep{birkhead2025algorithms}. In a market setting, a noisy signal exacerbates information asymmetry, potentially leading to adverse selection where high-quality data is undervalued due to artificially high variance in the score.

Beyond differential privacy, secure multi-party computation (MPC) has also been used for privacy-preserving data valuation \citep{lindell2020secure}. For example, FedValue \citep{HanWangWuFang2025} applies MPC to compute Shapley values in vertical federated learning, where multiple parties collaboratively train a predictive model on tabular data and seek to attribute value to feature groups via a largely model-agnostic, information-theoretic metric. Instead of feature utility,  we focus on the utility of new data samples. \citet{HanWangWuFang2025} also focuses on tabular data while our proposed method can work with diverse data modalities as shown in \Cref{sec:exp}.

A complementary route to privacy preservation is to encrypt the data itself and perform computation directly on encrypted representations, rather than protecting outputs with noise as in differential privacy or relying on interactive secure protocols among multiple parties as in MPC. This direction is especially attractive in our setting because the buyer and seller do not fully trust each other, yet both need a verifiable utility signal before any data transfer occurs.
Homomorphic Encryption (HE) is such a suitable family of cryptographic techniques that permits computation over ciphertexts while preserving the semantics of the corresponding plaintext operations after decryption. Within this family, Fully Homomorphic Encryption (FHE) refers to schemes that support sufficiently rich compositions of additions and multiplications (and thus general computations) on encrypted data, making them suitable for outsourced computation without revealing inputs \citep{marcolla2022survey}. 

To support valuation at scale, we instantiate our protocol with the CKKS scheme, an FHE scheme for approximate arithmetic that is particularly well-suited to real-valued vector operations such as gradient-based scoring in AI algorithms \citep{cheon2017homomorphic}. Compared to integer-oriented schemes such as BFV/BGV, CKKS avoids substantial quantization overhead in our setting because it natively supports approximate computations on packed real-valued vectors \citep{fan2012somewhat, brakerski2012fully, cheon2017homomorphic}. Existing CKKS-based systems mainly focus on privacy-preserving training or inference for fixed models, and, to the best of our knowledge, have not been used to operationalize in a data market setting.
Our work contributes by bringing modern cryptographic computation into the data-transaction stage itself, rather than using cryptography only to protect training or inference after data access is granted. We use FHE to make pre-purchase utility verification possible under mutual distrust.

\section{Secure Data Market}
\label{sec:method}

In this section, we formalize our secure data market and develop the core technical framework for privacy-preserving data valuation. In \Cref{sec:setup},  we define the market agents, private assets, and the value-privacy dilemma. In \Cref{sec:if-prelim}, we introduce the valuation metric based on the influence function and its tractable approximation for large models using gradient projections. Building on this, \Cref{sec:secure-market} describes the cryptographic primitives required for secure evaluation and assembles them into the Trustworthy Influence Protocol (TIP). Finally, in \Cref{sec:theoproperty}, we analyze the theoretical privacy guarantees for each agent in the system.

\subsection{Problem Setup}
\label{sec:setup}

We consider a data market in which companies that develop frontier AI models are the primary consumers of data. The key feature that distinguishes data from traditional commodities is that its value is model- and task-specific: the same data may be highly valuable to one model while being useless, or even harmful, to another. The fundamental economic friction in this market is the Arrow Information Paradox: a buyer cannot verify the value of data without inspecting it, but once having inspected it, the seller has effectively transferred the asset without compensation. Our goal is to formalize this setting and the constraints under which a secure market must operate.

Formally, the market consists of two primary agents:

\begin{itemize}

\item \textsc{Data Seller} ($\mathcal{S}$) holds a private candidate data point $z_s$ for sale, such as a specialized publisher with copyrighted textbooks or a hospital system with verified clinical data. The seller's goal is to monetize $z_s$ while maintaining the strict confidentiality of its content prior to a confirmed transaction.

\item \textsc{Data Buyer} ($\mathcal{B}$), a firm seeks to enhance a specific capability in an AI model $f_\theta$. 
The model is parameterized by a vector $\theta \in \Theta$ and trained on an existing dataset $D_{\text{train}}$. 
We assume $f_\theta$ is a publicly available architecture, such as Meta Llama 3 or DeepSeek-V3 \citep{grattafiori2024llama, liu2024deepseek}, where the current parameter vector $\hat{\theta}$ is known or reproducible by both parties
\footnote{We start with the open model assumption and relax this assumption in Section~\ref{extension:modelaccess}}.
For notation, let $D_{\text{train}}=\{z_i\}_{i=1}^{n}$ denote the buyer's existing training set. To verify utility, the buyer maintains a private evaluation set $D_{\text{eval}}=\{z_j^{(\mathrm{eval})}\}_{j=1}^{m}$ that serves as a proxy for the target task. Let $\ell(z;\theta)$ denote the per-sample loss (e.g., cross-entropy). We use $n$ and $m$ for sample sizes.
For instance, $D_{\text{eval}}$ may be a proprietary benchmark such as one analogous to GSM-8K \citep{cobbe2021gsm8k} for math reasoning, or one modeled after more discriminative reasoning-oriented evaluation suites such as MMLU-Pro \citep{wang2024mmlu}. 
In practice, AI models' performance on these evaluation benchmarks is often used as evidence of model improvement in a new model release.
For example, Google presents broad benchmark-based performance comparisons for Gemini 3.1 Pro (including evaluations such as MMMU-Pro among others) as part of its release narrative, and Anthropic similarly reports benchmark comparisons for Claude Sonnet 4.6 in its launch materials to substantiate model improvements \citep{Google2026_Gemini31Pro,Anthropic2026_ClaudeSonnet46}.

\end{itemize}

Given the buyer's training set $D_{\text{train}}=\{z_i\}_{i=1}^n$ and per-sample loss $\ell(z;\theta)$, we define the buyer's  sample-average training loss (empirical training risk) as:
$$
R_{\text{train}}(\theta) := \frac{1}{n} \sum_{z \in D_{\text{train}}} \ell(z; \theta). 
$$ 
The buyer's model is trained by minimizing this empirical training risk.
Let $\hat{\theta} \in \arg\min_{\theta \in \Theta} R_{\text{train}}(\theta)$ denote a minimizer of the empirical training risk for the buyer.
Accordingly, under first-order optimality conditions, we have $\nabla_\theta R_{\text{train}}(\hat{\theta})=0$. The buyer's performance on the target task is measured by the empirical evaluation risk:
$
R_{\text{eval}}(\theta) := \frac{1}{m} \sum_{j=1}^m \ell\bigl(z^{(\mathrm{eval})}_j; \theta\bigr).
$
Here, we assume the loss function $\ell$ is twice-differentiable and strictly convex in $\theta$. 
While deep neural networks are globally non-convex, we focus on the local geometry around the converged parameters $\hat{\theta}$. 
Since $\hat{\theta}$ is a local minimizer of the training risk, the Hessian matrix $H_{\hat{\theta}} := \nabla_\theta^2 R_{\text{train}}(\hat{\theta})$ is positive semi-definite by definition. In the context of influence functions, it is standard practice to add a small damping term $\lambda I$ to the Hessian to ensure it is positive definite and invertible \citep{choe2024your}. Therefore, the assumption of local strict convexity is technically justified within the trust region of the first-order approximation and does not limit the applicability of our framework to deep learning models.

If the buyer were to acquire $z_s$ and incorporate it into the training process either by retraining or fine-tuning with $z_s$ added to the training data, the resulting parameters would change from $\hat{\theta}$ to a new vector, denoted by $\hat{\theta}_{z_s}$, determined by the buyer’s fixed training procedure. The buyer’s objective is to purchase $z_s$ if and only if this retraining would reduce the evaluation loss on $D_{\text{eval}}$. The central challenge is to quantify this potential improvement before $z_s$ is revealed in plaintext.

The difficulty is that neither party is willing to reveal the information needed to compute this counterfactual change directly. The buyer’s private assets include the training dataset $D_{\text{train}}$ and the evaluation set $D_{\text{eval}}$; the seller’s private asset is the raw content of $z_s$. If the seller knows what task the buyer will be trained on, they will be able to strategically choose data or create data that will sell to the buyer; on the other hand, the buyer will free-ride from the seller if they know the value of the data prior to the transaction.
Thus, neither is willing to disclose their data to the other or to any external computing service without a security guarantee. 
The problem we study is therefore the following: design a practical mechanism that allows the buyer to obtain an accurate, quantitative assessment of how acquiring $z_s$ would affect the buyer’s evaluation risk \(R_{\text{eval}}(\theta)\), under these privacy constraints and without retraining the model separately for each candidate point.

To formalize the economic incentives within this marketplace, we adopt two standard stylized assumptions. First, we posit that the buyer's objective of minimizing evaluation risk is aligned with profit maximization. As demonstrated by \citet{zhang2025fairshare}, data valuation is not merely a technical metric but an economic imperative; pricing mechanisms grounded in marginal contribution align incentives between builders and annotators, maximizing long-term buyer utility and ensuring market sustainability.
Second, regarding the transaction mechanism, we posit that compensation for data is determined by its utility. For the scope of this paper, we focus on the fundamental challenge of securely computing the utility signal $s(z_s)$ required to operationalize this function. While we adopt a direct relationship, such as linear scaling, in our experiments, our secure valuation framework provides the primitives to support arbitrarily complex pricing dynamics or auction mechanisms.

\subsection{Influence-Function-Based Data Valuation}
\label{sec:if-prelim}

In our market, the buyer’s objective is to quantify the marginal utility of a seller’s candidate data \(z_s\). Ideally, this utility is defined as the reduction in $R_{\text{eval}}(\theta)$ resulting from adding \(z_s\) to the training set. Since retraining the model for every candidate point is computationally infeasible, we adopt the influence-function framework \citep{koh2017understanding} to approximate this counterfactual outcome. Formally, the influence function serves as a first-order approximation of the change in model parameters and subsequently loss under an infinitesimal perturbation of the training distribution, which allows the buyer to predict the value of acquiring new data at a fraction of the computational cost.

We model the acquisition of a candidate point \(z_s\) by defining a perturbed objective where \(z_s\) is introduced with weight \(\epsilon\):

$$
R_{\text{train},\epsilon,z_s} (\theta) \;=\; R_{\text{train}}(\theta) + \epsilon \, \ell(z_s; \theta),
$$

where $\hat{\theta}_{\epsilon,z_s} = \arg \min_\theta R_{\text{train},\epsilon,z_s} (\theta)$ denote the minimizer of this perturbed risk.
The following lemma characterizes the sensitivity of the loss to this candidate data. We present the pointwise influence $\mathcal I(z_s,z_{\mathrm{eval}})$ for clarity; aggregating over an evaluation set is a straightforward extension discussed later.

\begin{lemma}[Influence of a Candidate Point]
\label{lem:if-loss}

The influence of an exogenous candidate point \(z_s\) on the loss of a specific evaluation point \(z_{\mathrm{eval}} \in D_{\mathrm{eval}}\) is given by:
    \begin{equation}
    \label{eq:if_loss}
        \mathcal{I}(z_s, z_{\mathrm{eval}})
        \;\equiv\;
        \left.\frac{d\, \ell \bigl(z_{\mathrm{eval}};\hat{\theta}_{\epsilon,z_s}\bigr)}{d\epsilon}\right|_{\epsilon=0}
        \;=\;
        -\,
        \nabla_\theta \ell\bigl(z_{\mathrm{eval}};\hat{\theta}\bigr)^\top
        H_{\hat{\theta}}^{-1}
        \nabla_\theta \ell(z_s;\hat{\theta}).
    \end{equation}
\end{lemma}

Here, $H_{\hat{\theta}} \overset{\mathrm{def}}{=} \nabla_{\theta}^{2} R_{\text{train}}(\hat{\theta})$ is the Hessian of the training risk and is positive definite by assumption, which guarantees the existence of $H_{\hat{\theta}}^{-1}$\footnote{In practice, we use the damped Hessian formulation to ensure it's positive definite:  $(H_{\hat\theta} + \lambda I)$.}.
Equation~\eqref{eq:if_loss} denotes the negated preconditioned inner product of the seller's gradient \(\nabla_\theta \ell(z_s)\) and the buyer's evaluation gradient \(\nabla_\theta \ell(z_{\mathrm{eval}})\), preconditioned by the inverse Hessian \(H_{\hat{\theta}}^{-1}\).
Notably, this formula admits a clean decomposition into a buyer-side quantity and a seller-side quantity, coupled only through an inner product. This bilinear structure is what enables the encrypted evaluation protocol developed in Section~\ref{sec:secure-market}: each party can prepare its component independently, and an untrusted broker can combine them under Homomorphic Encryption without observing either input.
A negative value of $\mathcal{I}(z_s,z_{\mathrm{eval}})$ indicates that upweighting $z_s$ would decrease the evaluation loss, and thus $z_s$ is beneficial for the buyer’s task.
The influence function targets evaluation loss rather than classification accuracy, because accuracy is a non-smooth, discrete metric, so small improvements in loss do not always flip predictions, making accuracy insensitive to marginal data contributions. Evaluation loss, by contrast, is continuous and differentiable, enabling the first-order approximation that underlies Eq.~\eqref{eq:if_loss}. In practice, data that systematically reduces evaluation loss also tends to improve accuracy by widening classification margins. 
Crucially, \(H_{\hat{\theta}}\) accounts for information redundancy. It acts as a metric tensor that down-weights gradient components in directions of high curvature and amplifies components in directions of low curvature. Thus, unlike simple cosine similarity, the influence function rewards data that provides novel structural information rather than merely redundant signals. Below we use $s(z_s):= \mathcal{I}(z_s,z_{\mathrm{eval}})$ for simplicity, to denote the IF score. 
In practice, a small damping term $\lambda I$ is added to $H_{\hat\theta}$ (as noted in Section~\ref{sec:setup}), which bounds this amplification at $1/\lambda$ and ensures numerical stability. 
Gradients in low-curvature directions also carry small variance ($\mathbb{E}[(e_i^\top g)^2]\approx\lambda_i$)  under the FIM approximation, so their net contribution to the influence score remains limited even after amplification. High-curvature directions, concentrating both gradient energy and learned model knowledge, therefore dominate the influence signal in practice, a property that justifies the low-rank gradient projection strategy detailed in Appendix~\ref{app:scalable}. 
Full derivation and the proof of Lemma~\ref{lem:if-loss} are provided in Appendix~\ref{app:lemma3-1}.

While Eq.~\eqref{eq:if_loss} is theoretically sound, it is computationally intractable for large models. 
For example, for an LLM with billions of parameters, the gradient vectors $\nabla_\theta \ell \bigl(z_{\mathrm{eval}};\hat{\theta}\bigr)$ have a size of billions. 
One cannot store billions of parameters' worth of gradients for every training example, and multiply these giant vectors. Second, even for smaller models, the Hessian 
\(H_{\hat{\theta}}\) becomes a $d \times d$ matrix, making storage and inversion impractical. 
To render this metric usable in a practical data market, we employ a gradient projection strategy inspired by recent advances in scalable influence estimation, such as TRAK \citep{park2023trak} and LoGra \citep{choe2024your}. Detailed implementation of the memory- and compute-efficient gradient projection algorithm for computing influence functions is provided in Appendix~\ref{app:scalable}.

\subsection{Secure Data Market Design}
\label{sec:secure-market}

As noted in Section~\ref{sec:related_work}, a key reason for adopting influence functions is their bilinear structure: the score $s(z_s) = -\langle \tilde{v}_{\mathrm{eval}},\, \tilde{g}(z_s) \rangle$ cleanly separates into a buyer-side vector and a seller-side vector, each computable independently. This structural separability is precisely what enables the encrypted protocol below.
The valuation metric in Section~\ref{sec:if-prelim} gives the buyer a way to score $z_s$, but only if the raw data $z_s$ is visible. However, the seller's core privacy constraint is precisely this: they cannot expose $z_s$ or its gradient to any party before a confirmed transaction. Our design goal is therefore to construct a mechanism that lets the buyer compute $s(z_s)$ from encrypted representations, without either party revealing their private inputs.

\subsubsection{Secure Influence Computation with Homomorphic Encryption}
\label{sec:he_ckks}

To address this privacy deadlock, we employ Homomorphic Encryption (HE). Homomorphic Encryption is a powerful cryptographic method that allows computations to be performed directly on encrypted data, yielding an encrypted result that, when decrypted, matches the result of operations performed on the plaintext. The result of the computation remains encrypted and can only be revealed by the holder of the secret key.

For calculating influence scores from neural network gradients, we require an HE scheme that supports approximate arithmetic on real numbers (floating-point operations), rather than just exact integers. We therefore utilize the CKKS scheme \citep{cheon2017homomorphic}. Unlike classical schemes (e.g., BFV) that operate on discrete integers, CKKS is designed to handle the continuous noise and precision inherent in machine learning models. We formalize the specific capability required for our market in the following lemma (technical details and proofs are provided in Appendix~\ref{app:ckks}):

\begin{lemma}[Approximate homomorphism of CKKS]
\label{lem:ckks-approx}
Let \(m_1,m_2 \in \mathbb{R}^{t}\) be real vectors encoded at the same scale and ciphertext level. There exist CKKS parameters such that:

\[
\begin{aligned}
    Dec_{\mathsf{sk}}(Enc(m_1) \oplus Enc(m_2)) &\approx m_1 + m_2 + \varepsilon_{\mathrm{add}},\\
    Dec_{\mathsf{sk}}(Enc(m_1) \otimes Enc(m_2)) &\approx m_1 \odot m_2 + \varepsilon_{\mathrm{mul}},
\end{aligned}
\]

\end{lemma}

The core mechanism of CKKS is to encode real numbers into integer polynomials by multiplying them with a large scaling factor $\Delta$, thereby treating the inevitable encryption noise not as a corruption, but as a manageable numerical error analogous to floating-point round-off. The error terms $\varepsilon_{\mathrm{add}}$ and $\varepsilon_{\mathrm{mul}}$ represent this noise, inherent to the underlying Ring Learning With Errors (RLWE) hardness assumption.
For addition operation, the noise grows linearly, which is mathematically negligible because the message is encoded with an exponential scaling factor $\Delta$ (e.g., $2^{40}$). The huge gap between the scaling factor and the noise bound ($\Delta \gg B$) ensures that the additive noise remains strictly confined to the insignificant decimal range of the data.

For multiplication, the noise initially grows by the magnitude of the message, because each message carries a certain level of noise after encryption. However, this does not compromise the computation accuracy for two reasons. 
First, since the input vectors $m$ are amplified by a massive scaling factor $\Delta$ (e.g., $2^{40}$) before encryption, this operation effectively "lifts" the significant information far above the noise floor, rendering the encryption noise equivalent to a microscopic perturbation relative to the amplified signal, similar to a rounding error\footnote{From a cryptographic perspective, the noise resides in the least significant bits (LSBs) of the plaintext polynomial.}.
Second, the CKKS scheme applies a rescaling operation immediately after multiplication. It reduces both the signal and the accumulated noise proportionally, thereby preserving the original signal-to-noise ratio and maintaining the relative error bound. 

The practical utility of this approximate arithmetic is evidenced by experimental results. For instance, evaluating a high-degree polynomial such as $x^{16}$ or even $x^{1024}$ preserves approximately 10 to 26 bits of precision in the output\footnote{$x^{16}$ requires four squaring operations because $16 = 2^4$.}. This corresponds to roughly 3 to 8 decimal digits of accuracy. In the context of machine learning, this precision is sufficient because algorithms like Stochastic Gradient Descent (SGD) are inherently robust to noise. 
In our design of a secure data market, this lemma provides the necessary primitive: it allows us to compute $s(z_s)$ in the ciphertext domain, which is the dot product between gradients. 
Leveraging CKKS performed on ciphertexts, ensuring that the gradients are never exposed, enables us to translate the necessary gradient matrix operations into secure homomorphic evaluations, thereby enabling the Trustworthy Influence Protocol described in the following section. More details and an example demonstrating how homomorphic encryption (HE) enables computations on encrypted data can be found in Appendix~\ref{app:ckks}.

\subsubsection{The Trustworthy Influence Protocol (TIP)}

Leveraging the secure arithmetic properties of CKKS, we propose the Trustworthy Influence Protocol (TIP), as shown in \Cref{fig:fhe}. The protocol involves three entities: the Buyer (who holds the model and evaluation set), the Seller (who holds the candidate data), and an untrusted Broker (who performs the computation).

Let $\tilde{g}(z_{\mathrm{eval}}) := \nabla_\theta \ell \bigl(z_{\mathrm{eval}};\hat{\theta}\bigr)$,
$\tilde{v}_{\mathrm{eval}} := \nabla_\theta \ell \bigl(z_{\mathrm{eval}};\hat{\theta}\bigr)^\top H_{\hat{\theta}}^{-1} $, 
and $\tilde{g}(z_s^{(i)}) := \nabla_\theta \ell(z_s^{(i)}; \hat{\theta})$. 
Assume $\tilde{g}(z_{\mathrm{eval}})$ and $\tilde{g}(z_s^{(i)})$ approximately follow the same distribution. The buyer’s influence-based utility for a seller's data $z_s$ can be written as
$$
s(z_s^{(i)}) := - \big\langle \tilde v_{\mathrm{eval}},\, \tilde{g}(z_s^{(i)}) \big\rangle,
$$
where $\tilde v_{\mathrm{eval}} \in \mathbb{R}^k$ is the buyer's preconditioned evaluation vector and $\tilde g(z_s)\in\mathbb{R}^k$ is the seller's projected gradient in the same low-dimensional space.\footnote{Throughout TIP, the notation $\nabla_\theta$ refers to the gradient restricted to the LoGra projection-layer parameters $\theta_{\mathrm{proj}} \subset \theta$, which is a jointly agreed-upon $k$-dimensional subspace (see Appendix~\ref{app:scalable}). Differentiating with respect to the full parameter vector $\theta$ is theoretically correct but cryptographically infeasible: encrypting a billion-dimensional gradient under CKKS is computationally prohibitive. The LoGra projection compresses each gradient to $\mathbb{R}^k$ while preserving the dominant influence signal, making both the computation of influence and the homomorphic evaluation tractable.}
A negative $s(z_s)$ indicates that upweighting $z_s$ is predicted to reduce the buyer’s evaluation loss. 
The protocol proceeds in four sequential phases:

\paragraph{Phase 1: Setup \& Encryption (Buyer).}

The buyer generates a public-secret key pair \((\mathsf{pk}_B, \mathsf{sk}_B)\). Using the public model $f_{\hat{\theta}}$ and their private evaluation data $z_{\mathrm{eval}} = (x_{\mathrm{eval}},\, y_{\mathrm{eval}})$, the buyer performs a forward pass to obtain the prediction $\hat{y}_{\mathrm{eval}} = f_{\hat{\theta}}(x_{\mathrm{eval}})$, computes the evaluation loss $\ell(z_{\mathrm{eval}};\hat{\theta})$ (e.g., cross-entropy between $\hat{y}_{\mathrm{eval}}$ and $y_{\mathrm{eval}}$), and backpropagates through the projection layer to obtain the raw gradient $\tilde{g}(z_{\mathrm{eval}}) \in \mathbb{R}^k$. The buyer then applies the inverse-Hessian preconditioner, approximated via KFAC on their private training data to form the preconditioned evaluation vector $\tilde{v}_{\mathrm{eval}} = \tilde{g}(z_{\mathrm{eval}})^\top H_{\hat{\theta}}^{-1} \in \mathbb{R}^k$. They encrypt this vector into a ciphertext $\mathsf{ct}_{\mathrm{eval}} = Enc_{\mathsf{pk}_B}(\tilde{v}_{\mathrm{eval}})$ and send it to the broker. 
In addition, the buyer generates an evaluation key $\mathsf{evk}$ and forwards it to the broker; $\mathsf{evk}$ enables ciphertext relinearization during the homomorphic multiplication in Phase~3 without revealing $\mathsf{sk}_B$ (see Appendix~\ref{app:ckks}).
Crucially, the buyer keeps $\mathsf{sk}_B$ secret, ensuring no one else can decrypt this vector.

\paragraph{Phase 2: Gradient Preparation (Seller).}

For each candidate data $z_s^{(i)} = (x_s^{(i)},\, y_s^{(i)})$, the seller computes the projected gradient using the publicly available model $f_{\hat{\theta}}$. Concretely, the seller performs a forward pass to obtain the model prediction $\hat{y}_s^{(i)} = f_{\hat{\theta}}(x_s^{(i)})$, computes the per-sample loss $\ell(z_s^{(i)};\hat{\theta})$ (e.g., cross-entropy between $\hat{y}_s^{(i)}$ and the true label $y_s^{(i)}$), and backpropagates to obtain the projected gradient $\tilde{g}(z_s^{(i)}) \in \mathbb{R}^k$.
They encrypt this vector into $\mathsf{ct}_{i} = Enc_{\mathsf{pk}_B}(\tilde{g}(z_s^{(i)}))$ using the buyer's public key and send these encrypted gradients to the broker. Note that the seller does not need the secret key to perform encryption.

\paragraph{Phase 3: Blind Scoring (Broker).}
The broker now possesses two sets of encrypted vectors but cannot read the content of either. Using the homomorphic properties of CKKS (Lemma~\ref{lem:ckks-approx}), the broker computes the encrypted dot product for each candidate data point:
\[
    \mathsf{ct}_{\mathrm{ip},i} = \textsc{RotateAndSum}(\mathsf{ct}_{\mathrm{eval}} \otimes \mathsf{ct}_{i}).
\]
This operation essentially performs element-wise multiplication and summation in the encrypted domain. The resulting ciphertexts \(\{\mathsf{ct}_{\mathrm{ip},i}\}_{i=1}^N\) are sent back to the buyer.

\paragraph{Phase 4: Decryption \& Valuation (Buyer).}
The buyer uses their secret key \(\mathsf{sk}_B\) to decrypt the scores:
\[
    \hat{s}(z_s^{(i)}) \approx Dec_{\mathsf{sk}_B}(\mathsf{ct}_{\mathrm{ip},i}).
\]
The buyer now has the plaintext influence scores, which quantify the marginal utility of the seller's data, while the seller and broker have learned nothing about the buyer's evaluation set. The complete algorithmic procedure is detailed in \Cref{fig:fhe}.

\begin{algorithm}
\caption{TIP: Trustworthy Influence Protocol}
\label{alg:secure_influence}
\begin{algorithmic}[1] 
\Require Seller’s projected gradients $\tilde{g}(z_s^{(i)})$
\Require Buyer’s preconditioned eval vector $\tilde{v}_{\mathrm{eval}}$
\Ensure Buyer-only plaintext scores $\{\hat{s}(z_s^{(i)})\}_{i=1}^N$
\State $(\mathsf{pk}_B,\mathsf{sk}_B) \gets \textsc{CKKS.Setup}()$ \Comment{Buyer}
\State $\mathsf{ct}_{\mathrm{eval}} \gets Enc_{\mathsf{pk}_B}(\tilde{v}_{\mathrm{eval}})$ \Comment{Buyer}
\State Buyer sends $\mathsf{ct}_{\mathrm{eval}}$ to broker
\For{$i = 1$ to $N$}
    \State $\mathsf{ct}_i \gets Enc_{\mathsf{pk}_B}(\tilde{g}(z_s^{(i)}))$ \Comment{Seller}
    \State Seller sends $\mathsf{ct}_i$ to broker
    \State $\mathsf{ct}_{\mathrm{prod},i} \gets \mathsf{ct}_{\mathrm{eval}} \otimes \mathsf{ct}_i$ \Comment{Broker}
    \State $\mathsf{ct}_{\mathrm{ip},i} \gets \textsc{RotateAndSum}(\mathsf{ct}_{\mathrm{prod},i})$ \Comment{Broker}
    \State Broker sends $\mathsf{ct}_{\mathrm{ip},i}$ to buyer
\EndFor
\For{$i = 1$ to $N$}
    \State $\hat{s}(z_s^{(i)}) \gets Dec_{\mathsf{sk}_B}(\mathsf{ct}_{\mathrm{ip},i})$ \Comment{Buyer}
\EndFor
\State \Return $\{\hat{s}(z_s^{(i)})\}_{i=1}^N$ 
\end{algorithmic}
\end{algorithm}

The Trustworthy Influence Protocol effectively resolves the information asymmetry inherent in data markets by decoupling valuation from disclosure. By executing the influence computation in the encrypted domain, TIP enables the buyer to verify the marginal utility of the seller's assets ex ante without exposing their own proprietary model gradients. The cryptographic security of CKKS ensures that the broker and seller 
learn nothing about the buyer's evaluation set (beyond the scale of the gradients), while the buyer learns nothing about the seller's data content beyond the final scalar score.

\noindent \textbf{Dataset Acquisition:} While TIP describes the valuation of candidate data, it naturally extends to dataset evaluation. A common approach in the literature for valuing a subset $S$ is to assume that the collective influence is approximately additive, i.e., the sum of the individual influences of the data points in $S$ \citep{koh2019accuracy, broderick2020automatic}. Under this standard assumption, the buyer can value a dataset $S$ by simply summing the secure scores obtained via TIP: $s(S) \approx \sum_{z \in S} s(z)$. \citet{hu2024most} proposed an adaptive greedy heuristic, which accounts for the interaction between data points. We provide more discussion of these conditions in Appendix~\ref{app:additive_theorem}.

\subsubsection{Example of Secure Influence Score Computation}

To demonstrate the Trustworthy Influence Protocol (TIP) in action, we trace the computation of the influence score for a single candidate datum. We consider a simplified scenario in which the projected gradient dimension is $k=2$ and the scaling factor is $\Delta = 10^4$ to preserve 4 decimal places of precision. This setup explicitly demonstrates the trade-off between cryptographic efficiency and numerical precision (truncation error) inherent to the CKKS scheme. The agents are the Buyer ($\mathcal{B}$), the Seller ($\mathcal{S}$), and the Broker ($\mathcal{R}$).

\begin{enumerate}
    \item \textbf{Setup and Encryption (Buyer)}
    
    The Buyer generates a key pair $(\mathsf{pk}_B, \mathsf{sk}_B)$ and an evaluation key $\mathsf{evk}_B$. $\mathcal{B}$ keeps $\mathsf{sk}_B$ secret and forwards the evaluation key to the Broker, and forwards $(\mathsf{pk}_B)$ to the Seller.
    
    The Buyer first calculates their preconditioned evaluation vector $\tilde{v}_{\mathrm{eval}} = \nabla \ell(z_{eval})^\top H^{-1}$. The Buyer estimates the Inverse Hessian $H^{-1}$ using a scalable approximation method (e.g., KFAC) based solely on their private training data, capturing the curvature of the loss landscape without exposing the test dataset itself.
    Let the computed vector be:
    \[
    \tilde{v}_{\mathrm{eval}} = [0.54321, \; 1.23456].
    \]
    The Buyer encodes this vector using $\Delta = 10^4$. Note that since the input has 5 decimal places but the scale is only $10^4$, precision loss will occur during encoding:
    \[
    m_{\mathrm{eval}} = \lfloor \tilde{v}_{\mathrm{eval}} \cdot 10^4 \rceil = [5432, \; 12346].
    \]
    Observe that $0.54321$ becomes $5432.1 \to 5432$, losing the last digit.
    $\mathcal{B}$ encrypts this message into $\mathsf{ct}_{\mathrm{eval}}$ using $\mathsf{pk}_B$ and sends it to the Broker.

    \item \textbf{Gradient Preparation (Seller)}
    
    Suppose the Seller holds a candidate datum $z_s$ with a projected gradient:
    \[
    \tilde{g}(z_s) = [2.00000, \; -0.50000],
    \]
    where the ideal alignment is:
    \[
    \text{Alignment} = (0.54321 \times 2.0) + (1.23456 \times -0.5) = 1.08642 - 0.61728 = 0.46914.
    \]
    The expected influence score is $s(z_s) = - \text{Alignment} = -0.46914$.
    
    The Seller encodes the gradient at the same scale $\Delta = 10^4$:
    \[
    m_{\mathrm{grad}} = \lfloor \tilde{g}(z_s) \cdot 10^4 \rceil = [20000, \; -5000].
    \]
    Using $\mathsf{pk}_B$, the Seller encrypts $\mathsf{ct}_{\mathrm{grad}} \leftarrow \mathsf{Enc}_{\mathsf{pk}_B}(m_{\mathrm{grad}})$ and sends it to the Broker.

    \item \textbf{Blind Scoring (Broker)}
    
    The Broker holds two ciphertexts $\mathsf{ct}_{\mathrm{eval}}$ and $\mathsf{ct}_{\mathrm{grad}}$. To compute the dot product, the Broker performs homomorphic multiplication followed by summation.\footnote{For clarity of exposition, we omit encryption notations in the following derivations; all computations are nevertheless carried out in the ciphertext domain, ensuring that the broker cannot access any information about the underlying data.}
    
    \begin{itemize}
        \item \textbf{Element-wise Multiplication:}
        The Broker computes $\mathsf{ct}_{\mathrm{prod}} = \mathsf{ct}_{\mathrm{eval}} \otimes \mathsf{ct}_{\mathrm{grad}}$. In the plaintext domain, this corresponds to:
        \[
        m_{\mathrm{prod}}[0] = 5432 \times 20000 = 108,640,000
        \]
        \[
        m_{\mathrm{prod}}[1] = 12346 \times (-5000) = -61,730,000
        \]
        The message scale is now $\Delta^2 = 10^8$.
        
        \item \textbf{Rescaling:}
        The Broker applies $\mathsf{ct}_{\mathrm{res}} \leftarrow \mathsf{Rescale}(\mathsf{ct}_{\mathrm{prod}})$. This divides the underlying message by $\Delta = 10^4$:
        \[
        m_{\mathrm{res}}[0] = \lfloor 108,640,000 / 10000 \rceil = 10,864
        \]
        \[
        m_{\mathrm{res}}[1] = \lfloor -61,730,000 / 10000 \rceil = -6,173
        \]
        The scale is restored to $\Delta$, and the noise is reduced.
        
        \item \textbf{Summation (RotateAndSum):}
        The Broker sums the slots ($\textsc{RotateAndSum}$):
        \[
        m_{\mathrm{sum}} = 10,864 + (-6,173) = 4,691.
        \]
        Thus, the Broker sends the final ciphertext $\mathsf{ct}_{\mathrm{sum}}$ to the Buyer.
    \end{itemize}

    \item \textbf{Decryption and Valuation (Buyer)}
    
    The Buyer decrypts using $\mathsf{sk}_B$ to retrieve the integer $\mu \approx 4,691$.
    Decoding by $\Delta$:
    \[
    \text{Calculated Alignment} = 4,691 / 10^4 = 0.4691.
    \]
    Finally, applying the definition $s(z_s) = - \text{Alignment}$:
    \[
    \hat{s}(z_s) = -0.4691.
    \]
    Comparing the secure result ($-0.4691$) with the plaintext ground truth ($-0.46914$), we observe an absolute error of $4 \times 10^{-5}$. This deviation arises from rounding in Phases 1 and 3. However, the sign and magnitude are preserved, correctly signaling to the Buyer that this candidate data is beneficial.

\end{enumerate}

\subsection{Theoretical Property}
\label{sec:theoproperty}

In this section, we provide a formal security analysis of the Trustworthy Influence Protocol (TIP). We adopt a standard cryptographic threat model involving a malicious server (representing a compromised Broker or an eavesdropping adversary) that attempts to reconstruct the private inputs of the Buyer and Seller from the observed transcripts. 
To quantify the security risk, we rely on a property of many encryption algorithms: IND-CPA (Indistinguishability under Chosen-Plaintext Attack). Intuitively, IND-CPA guarantees that a ciphertext reveals no information about the underlying plaintext. This is often formalized as a game: suppose an adversary chooses two distinct messages, $m_0$ and $m_1$, and asks the system to encrypt one of them. If the adversary receives the ciphertext and cannot determine whether it corresponds to $m_0$ or $m_1$ with a probability significantly better than a random guess ($\frac{1}{2}$), the system is considered IND-CPA secure. This property implies that the encrypted data is computationally indistinguishable from random noise. 
Consequently, any attempt by a malicious party to reverse-engineer the raw data is equivalent to solving the underlying RLWE problem, which underpins the CKKS HE scheme \citep{cheon2017homomorphic}. 
The computational difficulty of this problem is quantified by the security parameter $\lambda$, a derived metric that depends on the polynomial modulus degree $N$ and the ciphertext modulus $Q$. Specifically in the CKKS setting, the hardness of the RLWE problem scales positively with the dimension $N$ and negatively with the modulus size $Q$. 
In our protocol design, we strictly adhere to the community standard \citep{HomomorphicEncryptionSecurityStandard} by selecting a sufficiently large $N$ relative to $Q$ (e.g., $N=8192$ for typical workloads) to guarantee a security level of $\lambda \ge 128$. 
This configuration ensures that the best known attack requires approximately $2^{128}$ fundamental operations. To put this in perspective, even with the most powerful supercomputers, such a computation would take billions of years, rendering brute-force attacks infeasible. Under this constraint, the adversary's ability to distinguish the ciphertext from random noise better than a random guess is mathematically proven to be negligible.
Based on this foundation, we establish three propositions quantifying the risk of data leakage for the Seller, the Buyer, and the Broker.

\subsubsection{Seller Data Confidentiality}

The Seller's primary asset is the proprietary data $z_s$, represented by the gradient vector $\mathbf{g}_s$. The security goal is to prevent any party from reconstructing $\mathbf{g}_s$.

\begin{proposition}[Seller's Data Leakage Risk]
Let $\mathcal{A}$ be a malicious adversary (Broker or Buyer) attempting to recover the raw gradient $\mathbf{g}_s$ from the protocol execution. Let $P_{\text{recover}}^{\text{Seller}}$ denote the probability of successful reconstruction. We have:
\begin{equation}
    P_{\text{recover}}^{\text{Seller}} \le 2^{-\lambda} + \epsilon_{\text{inference}}.
\end{equation}
\end{proposition}

The Seller's risk is mitigated through two layers:
\begin{enumerate}
    \item \textbf{Transmission Security (Against Broker):} The Seller transmits the encrypted gradient $\mathsf{ct}_{\text{grad}}$. Since the CKKS scheme satisfies IND-CPA (see Appendix~\ref{app:ckks}), the ciphertext is indistinguishable from random noise to the Broker, who lacks the secret key. The probability of breaking this encryption is bounded by the cryptographic margin $2^{-\lambda}$.
    \item \textbf{Reconstruction Security (Against Buyer):} The Buyer decrypts the scalar score $s = \langle \tilde{v}_{\text{eval}}, \mathbf{g}_s \rangle$. Even if the Buyer acts maliciously to infer $\mathbf{g}_s$, they face an under-determined linear system (1 equation, $k$ unknowns). The probability $\epsilon_{\text{inference}}$ of uniquely identifying the high-dimensional vector $\mathbf{g}_s$ from a single scalar approaches zero information-theoretically.
\end{enumerate}

\subsubsection{Buyer Task Confidentiality}

The Buyer's privacy concern is the confidentiality of the evaluation task, that is, specific test cases or strategic benchmarks, which are encoded in the vector $\tilde{v}_{\text{eval}}$.

\begin{proposition}[Buyer's Task Leakage Risk]
Let $\mathcal{A}$ be a malicious Seller or Broker observing the encrypted evaluation vector $\mathsf{ct}_{\text{eval}}$. Let $P_{\text{leak}}^{\text{Buyer}}$ be the probability that $\mathcal{A}$ successfully distinguishes the true task vector from a random vector. We have:
\begin{equation}
    P_{\text{leak}}^{\text{Buyer}} \le 2^{-\lambda}.
\end{equation}
\end{proposition}

In the TIP protocol, the Buyer generates the key pair and keeps the secret key $\mathsf{sk}_B$ private. The task vector is transmitted solely as a ciphertext $\mathsf{ct}_{\text{eval}}$. By the definition of IND-CPA, any adversary (including the Seller performing the multiplication) cannot distinguish the encryption of the true task $\tilde{v}_{\text{eval}}$ from the encryption of a dummy vector of zeros. Therefore, the information leakage regarding the nature of the downstream task is bounded by the negligible probability $2^{-\lambda}$.

\subsubsection{Broker Blindness}

The Broker is an intermediary entrusted with computation. The requirement is that the Broker remains "blind" to both the inputs and the computed results.

\begin{proposition}[Broker's View Blindness]
Let $P_{\text{dist}}$ be the probability that the Broker can distinguish the actual computation transcript from a simulated transcript of random numbers. We have:
\begin{equation}
    P_{\text{dist}} \le 2^{-\lambda}.
\end{equation}
\end{proposition}

The Broker operates exclusively in the ciphertext domain. All inputs ($\mathsf{ct}_{\text{eval}}, \mathsf{ct}_{\text{grad}}$), intermediate states, and outputs are valid ciphertexts. Due to the semantic security of the underlying scheme, these ciphertexts appear as uniform random strings to any party without the decryption key. Consequently, the Broker learns zero information about the data distribution, the model parameters, or the final valuation score.

Consequently, in the real world, the probability that a malicious server can successfully compromise the confidentiality of the transaction, whether by recovering the seller's raw data or the buyer's evaluation logic, is astronomically low, providing a robust trust foundation for sensitive data exchange. To the best of our knowledge, our study is among the first to design a cryptographically secure valuation framework that operationalizes gradient-based influence functions directly on encrypted data, effectively resolving Arrow's Information Paradox without relying on a trusted third party.

\section{Experiments}
\label{sec:exp}

A central requirement for a secure data valuation framework is that it must remain both faithful to the underlying influence-function calculation and scalable across modern deep architectures. In practice, this means the encrypted influence scores must closely match their plaintext counterparts numerically, so that a data buyer or regulator would make identical allocation or compensation decisions even without accessing sensitive data. Second, the computational procedures for encrypted gradient logging and homomorphic influence computation must remain tractable as models grow from small MLPs to billion-parameter LLMs. This section evaluates fidelity and scalability across three representative tasks of increasing complexity: 

\begin{enumerate}
    \item \textbf{Image classification with an MLP on MNIST}: We employ a standard Multi-Layer Perceptron (MLP) trained on the Modified National Institute of Standards and Technology (MNIST) dataset \citep{6296535}. This foundational computer vision benchmark consists of 70,000 grayscale images of handwritten digits from 0 to 9. Due to its simplicity, this setup serves as a controlled baseline, allowing us to compute exact influence scores without approximation, thereby strictly verifying the numerical precision of the cryptographic operations.

    \item \textbf{Sentiment analysis with BERT}: We fine-tune the BERT-base (Bidirectional Encoder Representations from Transformers) architecture on the Stanford Sentiment Treebank (SST-2) dataset \citep{socher-etal-2013-recursive}. This dataset is a core component of the GLUE benchmark \citep{wang2018glue}, containing movie reviews annotated with binary sentiment labels (positive/negative). This experiment tests the protocol's compatibility with modern discriminative NLP workflows and evaluates the efficacy of KFAC preconditioners in handling the complex attention layers of Transformer models.

    \item \textbf{Next-token prediction with GPT-2}: We apply the framework to a Generative Pre-trained Transformer 2 (GPT-2) \citep{radford2019language} model trained on the WikiText-2 dataset \citep{merity2016pointer}. This corpus comprises over 2 million tokens extracted from rigorous, high-quality Wikipedia articles. As a causal language modeling task, this setting represents the frontier of Generative AI, assessing the framework's scalability to the autoregressive mechanisms and high-dimensional parameter spaces characteristic of Large Language Models (LLMs).
\end{enumerate}

Throughout these experiments, we preserve model expressiveness while compressing the influence-relevant gradients into a homomorphically tractable representation. In all models, we insert lightweight LoGra adapters so that gradients flowing through them serve as a low-dimensional projection of each data example. Only these compressed gradients are ever encrypted, shared, or processed under homomorphic operations. This ensures strict privacy while maintaining a meaningful directional approximation of the full gradient field used by classical influence functions. The broker then computes the required influence expression via CKKS multiplication, rotation, and slot summation, without learning anything about the original data or the underlying model parameters.

We then quantify the computational overhead introduced by CKKS-based encryption and inner-product evaluation, showing that these results validate the cryptographic layer's preservation of the influence-function-based valuation and its practicality for downstream data market simulations. Furthermore, we found that the per-sample cost depends primarily on the projected gradient dimension rather than the underlying model size.

\subsection{Approximation Consistency}
\label{sec:validity}

To rigorously validate that our encrypted valuation signal ($s_{\text{IF}}$) serves as a reliable proxy for realized model improvement, we conducted an experiment using the MNIST dataset on a standard MLP to verify whether the protocol can correctly identify a seller offering the data that the buyer lacks, versus redundant data, solely based on encrypted gradients.

We construct a single-digit market scenario characterized by a deliberate distributional shift between the buyer's training data and the evaluation target. The buyer possesses a private training set $D_{\text{train}}$ consisting exclusively of digits $\{1, 2\}$. The buyer's model (a 3-class MLP) is trained to converge on this subset. Crucially, the buyer has never seen digit $3$. Thus, the buyer seeks to improve performance on digit $3$, where the evaluation set $D_{\text{eval}}$ consists solely of images of digit $3$. There are three sellers offering distinct partitions of data:
\begin{itemize}
    \item \textbf{Seller A:} Offers data containing only digit $1$ (Redundant).
    \item \textbf{Seller B:} Offers data containing only digit $2$ (Redundant).
    \item \textbf{Seller C:} Offers data containing only digit $3$ (Novel).
\end{itemize}

Based on the distribution of the sellers' data, it is clear that seller C's data is most useful to buyers.
However, the buyer doesn't know the distribution of data across sellers in advance. 
Then we use TIP to perform the secure data trading. For each seller, we compute two metrics: FHE-IF Score ($s_{\text{IF}}$), and realized utility $ \Delta \mathcal{L}$, which is the ground-truth change in evaluation loss after acquiring the seller's data and performing a one-shot adaptation of the model. A negative $\Delta$ Loss indicates actual performance improvement.
The results of approximation consistency over 100 Replicates are presented in \Cref{tab:appro_concistency}.

\begin{table}[ht]
\centering
\small
\renewcommand{\arraystretch}{1.2}
\begin{threeparttable}

\caption{\textbf{Approximation Consistency in Single-Digit Market over 100 Replicates.}}

\begin{tabular*}{\textwidth}{@{\extracolsep{\fill}}lccc}
\toprule
\textbf{Method} 
& $\mathbf{\mathcal{L}}$ 
& $\mathbf{\Delta \mathcal{L}}$ 
& \textbf{FHE-IF score} $(\times 10^{-3})$ \\
\midrule

Baseline 
& 15.72 (0.28) 
& -- 
& -- \\

sellerA (\textit{1}) 
& 14.02 (0.21) 
& $-1.70^{***}$ (0.09) 
& -0.19 (0.36) \\

sellerB (\textit{2}) 
& 15.98 (0.29) 
& $0.27^{***}$ (0.01) 
& 0.11 (0.17) \\

sellerC (\textit{3}) 
& 0.01 (0.000) 
& $-15.72^{***}$ (0.281) 
& $-386.78^{***}$ (11.921) \\

\bottomrule
\end{tabular*}

\begin{tablenotes}[flushleft]
\footnotesize
\item \textit{Note.} Entries are means across completed replicates with standard errors in parentheses.
Negative influence scores indicate samples predicted to improve model performance.\\
$^{***}p<0.01$.
\end{tablenotes}

\end{threeparttable}
\label{tab:appro_concistency}

\end{table}

The baseline model, having never seen the digit $3$, exhibits a high initial loss ($\mathcal{L} \approx
  15.57$). 
Seller C, who provides the missing digit $3$, is theoretically the most valuable contributor. This is reflected in the realized utility, where acquiring Seller C's data almost eliminates the loss ($\Delta \mathcal{L} \approx -15.57$). 
Correspondingly, our secure protocol assigns Seller C a massive negative influence score ($s_{\text{IF}} \approx -386.78$). The magnitude of this score is two orders of magnitude larger than that of the other sellers, providing a decisive signal for the buyer to purchase from Seller C.

Moreover, Sellers A and B offer data with digits $1$ and $2$ that the buyer already possesses. The realized utility of acquiring such data is marginal (Seller A reduces loss slightly, Seller B increases it slightly due to distribution shift). Crucially, the FHE-IF scores for these sellers are close to zero relative to Seller C. This confirms that the Hessian-based preconditioning ($H^{-1}$) effectively down-weights redundant gradients, preventing the buyer from overvaluing data that offer no new information.

In conclusion, 
the FHE-IF score correctly identifies Seller C as the only high-utility provider. The magnitude and sign of the influence score align perfectly with the realized reduction in loss ($\Delta$ Loss) after adaptation.
The strong alignment between the ranked influence scores and the realized loss reductions confirms that FHE-IF serves as a valid proxy for model improvement, enabling buyers to distinguish between essential and redundant data without accessing the raw samples.

\subsection{Scalability}
\label{sec:scalability}

While the previous section established the approximation consistency of our influence-based metric, its practical adoption hinges on two critical engineering constraints: cryptographic fidelity and computational scalability. First, the homomorphic evaluation must reproduce plaintext influence scores with high numerical precision; otherwise, the privacy-preserving layer would introduce noise that distorts market signals. Second, the protocol must remain computationally tractable as models scale from simple networks to billion-parameter LLMs. In this subsection, we rigorously stress-test the Trustworthy Influence Protocol (TIP) across a spectrum of model architectures. To ensure feasibility, we leverage Low-Rank Gradient Projection (LoGra) to compress high-dimensional gradients into compact, encrypted vectors. We examine the framework's performance on three representative tasks of increasing complexity: MLP structure on the MNIST dataset, BERT on the SST-2 dataset, and GPT-2 model on the WikiText-2 dataset.

\subsubsection{Multilayer Perceptron with MNIST Dataset}

The MNIST experiment evaluates the basic validity of our approach in a controlled, low-complexity environment where plaintext influence is easy to compute exactly. Here, LoGra adapters reduce the MLP’s full gradient from over half a million parameters to a compressed vector of dimension 8,292. These vectors are encrypted under CKKS and passed to an untrusted broker, who computes the influence score via a homomorphic inner product. This setup allows us to verify whether the encrypted pipeline introduces numerical distortions or stability issues before scaling to larger models.

\subsubsection{BERT on SST-2 Sentiment Classification}

To assess scalability to modern Transformer architectures, we next apply the framework to BERT-base. The gradient dimensionality of BERT is far too large for direct encryption, so we attach LoGra adapters with rank 8 to all attention and feed-forward blocks, yielding compact 4,676-dimensional projected gradients. Unlike the MNIST case, here the influence computation involves a KFAC-based inverse-Hessian preconditioner, which ensures that the low-dimensional projection keeps the data that actually matter for the model's learning. Testing encrypted influence under this setting validates the feasibility of preconditioned influence functions within the encrypted protocol.

\subsubsection{Generative LLM: GPT-2 on WikiText-2}

We extend our evaluation to a large generative language model, GPT-2, on a next-token prediction task over the WikiText-2 corpus. GPT-2 presents unique challenges: gradients span tens of millions of parameters, and autoregressive architectures require deeper stacks of attention–MLP layers. We therefore adopt a surgical LoGra placement strategy, applying low-rank adapters (rank 4) only to the feed-forward sublayers that dominate directional gradient flow during next-token prediction. This insight reduces the per-example gradient to a remarkably compact 384-dimensional vector, which is small enough for fast encrypted operations yet rich enough to reproduce plaintext influence values with near-perfect fidelity.

\subsubsection{Results}
\label{results}

We evaluate our encrypted data evaluation framework on the above three representative tasks of increasing scale and complexity.  For each task, we report: 
\begin{enumerate}
    \item 
    \textbf{Model performance} (e.g., classification accuracy or perplexity) under normal, unencrypted training. 
    \item \textbf{Fidelity of encrypted influence scores} is measured both by absolute error versus the plaintext baseline and by Pearson correlation.
    \item \textbf{Runtime overhead} introduced by our homomorphic steps, specifically the time to encrypt and decrypt the logged gradients, and the time to compute the influence function under CKKS.
\end{enumerate}

\begin{table}[ht]
\centering
\small
\renewcommand{\arraystretch}{1.2}

\caption{\textbf{Experiment Results.} The proposed method achieves near-perfect
data utility computation with a computation time independent of model size.}
\label{tab:fhe_result}

\begin{tabular*}{\textwidth}{@{\extracolsep{\fill}}lcccc}
\toprule
\textbf{Dataset/Model}  
& \textbf{Pearson Correlation} 
& \textbf{Mean Absolute Error} 
& \textbf{Extra Running Time} 
& \textbf{Time Per Sample} \\
\midrule

MNIST / MLP     
& 1.00 
& $2.16\times10^{-5}$ 
& 890.19 s 
& 0.1483 \\

SST-2 / BERT     
& 0.97 
& $1.84\times10^{-5}$ 
& 10016.07 s 
& 0.1495 \\

WikiText-2 / GPT-2       
& 1.00 
& $1.12\times10^{-5}$ 
& 3.1 s     
& 0.1476 \\

\bottomrule
\end{tabular*}

\end{table}

Across all three tasks, encrypted influence scores closely match plaintext ones. High fidelity ensures FHE-based valuations match plaintext decisions. These findings confirm two central claims, fidelity and scalability. First, secure influence computation is precise enough to make identical data valuation decisions as the plaintext method. Second, once gradients are projected through LoGra, the extra computational cost depends chiefly on the dimensionality of the encrypted vector rather than the underlying model size, making encrypted influence feasible even for billion-parameter LLMs.

Also, the per-sample runtime observed ($\approx$ 0.15 seconds) suggests that the additional FHE overhead is linear in the number of samples and stable across tasks; budgeting teams can accurately forecast the computing expense for any scale, from a small 21‐sample probe to tens of thousands of data records. Organizations can therefore estimate both the utility of acquiring new data and the computational budget required to evaluate it, while ensuring that sensitive data never leaves encrypted form.

\subsection{Case Studies}
\label{sec:casestudy}

In this section, we instantiate our framework in two stylized data markets to study its practical implications. We first construct a healthcare data market using inpatient case-mix records, where we can explicitly compare encrypted influence scores against ground-truth utility obtained from controlled model retraining, thereby assessing the approximation consistency of our valuation signal. We then simulate a book data market for generative AI training, using an LLM fine-tuned on a reasoning benchmark to quantify the heterogeneous marginal utility of individual books. Together, these simulations with real data illustrate how secure influence-based valuation can support privacy-preserving data collaboration or trading in regulated domains and reveal extreme heterogeneity that challenges flat-rate compensation schemes in copyright-driven markets.

To validate the proposed framework in a domain characterized by high data heterogeneity and stringent privacy regulations, we simulate a data market using inpatient case-mix records from the Maryland Health Services Cost Review Commission (HSCRC\footnote{https://hscrc.maryland.gov}). The healthcare sector presents a paradigmatic example of the ``value-privacy dilemma.'' While hospitals possess vast repositories of patient data that could theoretically enhance peer institutions' predictive models, these datasets are siloed due to privacy risks (the Health Insurance Portability and Accountability Act of 1996 (HIPAA)\footnote{https://www.hhs.gov/hipaa}) and competitive concerns. Furthermore, because hospitals differ substantially in patient demographics, comorbidity profiles, and care practices, the external validity of any single dataset is uncertain. A blind acquisition of external data creates significant adverse selection risks, where a buyer might purchase data that fails to improve their model's performance.

We operationalize the market simulation using the HSCRC dataset, which contains the complete 1.29 million inpatient discharge records across 54 hospitals in Maryland from 2016-2025. Each record includes rich clinical features, including varying diagnosis codes\footnote{We followed the International Classification of Diseases, 10th version (ICD-10: \url{https://www.icd10data.com/ICD10CM/Codes})}, utilization metrics (length of stay), and severity indices. We employ a feature encoder that maps these records into a 5,273-dimensional sparse vector space.

The market consists of two distinct agents, defined as follows:

\begin{itemize}
    \item \textbf{The Buyer (Model Owner):} A hospital seeking to improve a binary classifier (Multilayer Perceptron) that predicts inpatient revisit risk (30-day readmission). The Buyer possesses a local private dataset of $N_{train}=2,000$ labeled encounters and a held-out evaluation set of $N_{eval}=2,000$. The Buyer's objective is to identify external data that minimizes the cross-entropy loss on their local evaluation set.
    \item \textbf{The Sellers (Data Owners):} Peer hospitals (up to $N=10$ per replication), each offering a bundle of $N_{seller}=2,000$ encrypted patient records.\footnote{Sellers with fewer than $N_{seller}=2,000$ available records are excluded from that replication, resulting in a variable number of effective sellers per replication (range: 4--10; mean: $\approx 7.6$).} Sellers cannot reveal raw data due to privacy constraints.
\end{itemize}

We conducted 50 independent market replications. In each replication, we randomly assign one hospital as the Buyer and ten others as Sellers. The evaluation proceeds in three stages:

\begin{enumerate}
    \item \textbf{Baseline Training:} The Buyer trains a baseline MLP on their local data until convergence.
    \item \textbf{Ex ante Encrypted Data Valuation:} Before any data transfer occurs, the Buyer and Sellers execute our secure valuation protocol. To assess the value of model access, we evaluate three valuation signals that represent a progression of model information requirements:

    \begin{itemize}
          \item \textbf{Data-Cosine.}
          This variant requires no access to the buyer's model and is applicable
          in settings where the buyer deploys a closed or proprietary system.
          It measures feature-space alignment between the seller's encoded record
          $z_i$ and the buyer's evaluation record $z_{\mathrm{eval}}$:
          \begin{equation}
              s_{\mathrm{DataCos}}(z_i) = \frac{z_i^\top z_{\mathrm{eval}}}
              {\|z_i\| \, \|z_{\mathrm{eval}}\|}.
          \end{equation}
          Because it requires only the raw feature vectors, this signal is
          computationally trivial but contains no information about the
          buyer's learning objective or the current state of the model.

          \item \textbf{Gradient-Cosine.}
          This variant incorporates first-order model information. Both the seller and the buyer compute projected gradients via the publicly available model $f_{\hat\theta}$, and the valuation score is the cosine similarity of their gradient vectors: 
          \begin{equation}
          s_{\mathrm{GradCos}}(z_i) = \frac{\tilde{g}(z_i)^\top \tilde{g}(z_{\mathrm{eval}})}{\|\tilde{g}(z_i)\| \, \|\tilde{g}(z_{\mathrm{eval}})\|}, 
          \end{equation} 
          where $\tilde{g}(z) = \nabla_\theta \ell(z; \hat{\theta}) \in \mathbb{R}^k$ is the projected gradient. Compared to Data-Cosine, this signal captures whether the seller's data points in the same direction as the buyer's current learning need. However, it treats all gradient directions as equally important, implicitly assuming a flat (isotropic) loss landscape, and therefore ignores whether a given direction is already well-mastered by the model or remains unexplored.

          \item \textbf{FHE-IF via TIP.}
          The full Trustworthy Influence Protocol adds second-order curvature information via the inverse Hessian $H_{\hat\theta}^{-1}$. This reweights gradient directions by the loss landscape geometry: directions of low curvature (underexplored, high marginal utility) are amplified, while directions of high curvature (already mastered, redundant) are suppressed. The result is a curvature-aware marginal utility estimate that directly approximates the counterfactual reduction in evaluation loss if the seller's data were added to training (Lemma~\ref{lem:if-loss}).

          \item \textbf{Random (Null Hypothesis):} As a randomization control, seller scores are drawn i.i.d.\ from a standard normal distribution, independent of the data and the model.
          \begin{equation}
              s_{\mathrm{Rand}}(z_i) \sim \mathcal{N}(0, 1).
          \end{equation}
          Scores are regenerated independently for each replication. This control tests whether any observed correlation is an artifact of the experimental pipeline itself; failure of FHE-IF and the cosine variants statistically outperforming this baseline would indicate that the market contains no exploitable valuation structure.
          
      \end{itemize}

    \item \textbf{Ex Post Ground Truth Realization:} To determine the true utility of the data, we simulate a counterfactual purchase. The Buyer acquires the Seller's data and performs a controlled fine-tuning (adaptation) of the model's classification head for one epoch. The realized benefit is defined as the reduction in evaluation loss.
\end{enumerate}

Results are reported at \Cref{tab:healthcare}. We report the mean absolute correlations between predicted valuation scores and realized utility across 50 market replications (384 buyer--seller pairs). The three tiers represent increasing levels of model-information access: Data-Cosine (feature similarity, no model), Gradient-Cosine (first-order gradient direction), and FHE-IF (second-order, Hessian-weighted influence estimation via TIP). The Lift column reports the mean paired difference (FHE-IF $-$ Gradient-Cosine) within each replication to control for market heterogeneity, with 95\% confidence intervals.

\begin{table}[htbp]
    \centering
    \small
    \renewcommand{\arraystretch}{1.2}
    
\begin{threeparttable}

\caption{\textbf{Healthcare Data Market: Valuation Accuracy Across Valuation Tiers.}}
\label{tab:healthcare}

\begin{tabular*}{\textwidth}{@{\extracolsep{\fill}}lccccc}
\toprule
\textbf{Metric}
& \textbf{FHE-IF}
& \textbf{Grad-Cosine}
& \textbf{Data-Cosine}
& \textbf{Random}
& \textbf{Lift (FHE-IF$-$GradCos)} \\
\midrule
$|$Pearson$|$ ($r$)
& 0.959
& 0.913
& 0.285
& 0.354
& $+0.048^{***}$ $[+0.019,\;+0.082]$ \\
$|$Spearman$|$ ($\rho$)
& 0.900
& 0.855
& 0.311
& 0.352
& $+0.047^{*}$ $[+0.001,\;+0.100]$ \\
\bottomrule
\end{tabular*}

\begin{tablenotes}[flushleft]
\footnotesize
\item Note: Values report mean absolute correlations between predicted
valuation scores and realized outcomes across experimental runs.
The Lift column reports the difference in performance between FHE-IF and
Grad-Cosine. 95\% confidence intervals are reported in brackets.
\item $^{***} p < 0.001$;\; $^{*} p < 0.05$.
\end{tablenotes}

\end{threeparttable}
\end{table}

The simulation yields 378 distinct buyer-seller evaluation pairs. \Cref{tab:healthcare} reports mean absolute correlations for each valuation tier.
The three-tier structure reveals a clear monotone progression. Data-Cosine, which uses only raw feature similarity without any model information, achieves the weakest correlation ($|r| = 0.285$), confirming that feature proximity alone provides only a coarse proxy for model utility.
Incorporating first-order model information yields a substantial gain: Gradient-Cosine improves Pearson correlation by $+0.628$ ($|r| = 0.913$), demonstrating that even without curvature weighting, model gradients encode far more task-relevant signal than raw features. The full FHE-IF protocol achieves the highest correlation ($|r| = 0.959$), with a statistically robust additional gain of $+0.048$ over Gradient-Cosine.
We note that model-agnostic valuation metrics based on information-theoretic quantities have been shown to perform effectively in federated data sharing settings where model internals are inaccessible by design \citep{HanWangWuFang2025}. Our three-tier comparison complements this finding: when the buyer operates a closed model, Data-Cosine and similar model-agnostic signals remain the only viable option but still provide a meaningful, if coarse, ranking of sellers. However, when the goal is to value data points (not data features) for a specific open AI algorithm, our results imply that incorporating model information yields progressively more precise utility estimates. 
The monotone progression across tiers thus suggests that the choice of valuation method should be matched to the level of model access available in a given market setting.

Although this second-order increment is smaller in absolute magnitude than the first-order step, it is systematic across all 50 replications, reflecting the benefit of Hessian preconditioning: by weighting gradient directions according to the buyer's specific loss curvature, FHE-IF distinguishes genuinely informative sellers from those whose data are directionally aligned but structurally redundant.

The exceptionally high Pearson correlation ($r \approx 0.96$) of the FHE-IF signal suggests that the metric effectively linearizes the relationship between the encrypted score and the realized clinical value. In a practical data marketplace, this property supports not just ordinal ranking, for instance, identifying the best hospital partner, but also linear pricing mechanisms. A buyer can confidentially allocate budget proportional to the encrypted influence score, knowing it serves as a precise proxy for the marginal reduction in clinical risk. This resolves the information asymmetry inherent in healthcare data markets without compromising patient privacy.

\subsubsection{Book Data Market Simulation}

While the healthcare simulation establishes the approximation consistency of our framework in a classical machine learning setting, the market for Generative AI training data presents distinct challenges regarding scalability and heterogeneity. In the healthcare context, model retraining (e.g., MLP) is computationally feasible, allowing us to verify that secure influence scores map directly to ground-truth utility. However, for LLMs, retraining on every potential data source to verify utility is computationally prohibitive. Consequently, current data marketplaces (e.g., AWS Data Exchange, Snowflake) typically rely on proxy signals such as data dictionaries, metadata descriptions, or small plaintext samples to facilitate pricing \citep{aws2024dataexchange}. These mechanisms assume an implicit homogeneity in data value or require privacy-compromising inspections.

This assumption of homogeneity is currently being challenged in the legal domain. The prevailing paradigm of indiscriminate scraping has precipitated high-profile copyright litigation, exemplified by the recent class-action settlement in \emph{Bartz v.\ Anthropic} (Case No. 3:24-cv-04760 (N.D. Cal.)) \citep{bartz2024anthropic}. Legal resolutions in this space often default to a ``flat-rate'' compensation model, estimating an average payout per work regardless of its semantic content \footnote{For context on flat-rate precedents, see the Google Book Search Settlement regarding per-book payouts: \url{https://techcrunch.com/2009/02/11/google-book-settlement-site-is-up-paying-authors-60-per-scanned-book/} and legal analysis of recent settlements: \url{https://legalblogs.wolterskluwer.com/copyright-blog/the-bartz-v-anthropic-settlement-understanding-americas-largest-copyright-settlement/}.}. From an economic perspective, such average-pricing mechanisms fail to account for the extreme heterogeneity of data utility; they treat a rigorous scientific textbook and a generic pamphlet as identical commodities. This lack of differentiation risks undercompensating the creators of high-utility data while incentivizing the production of low-quality content.

To resolve this friction, the data market requires a transition from broad settlements to ``meritocratic'' licensing. Managerially, this entails moving beyond the assumption that ``more data is better'' toward a protocol that fairly quantifies the marginal contribution of specific works. This allows model builders to justify premium compensation for high-value sources while filtering out data that actively degrades performance.

We operationalize this transition in a synthetic but realistic market. The Buyer is an OpenELM-270M language model optimizing for a specific downstream capability: the Multi-step Soft Reasoning (MuSR) benchmark \citep{sprague2023musr}. We specifically target the ``team allocation'' task, which requires satisfying complex logical constraints\footnote{The team allocation task asks the model to read a natural-language story, infer each person’s skills and interpersonal compatibility from soft clues, and assign people to tasks to maximize overall team efficiency.}. The sellers are 1,000 individual books drawn from the BookCorpusOpen dataset \citep{huggingface2024bookcorpus}. Using our secure protocol, the Buyer calculates the marginal utility of each book ex ante, enabling a licensing decision based on predicted performance rather than data volume.

\begin{figure}[ht] 
\centering 
    \includegraphics[width=\linewidth]{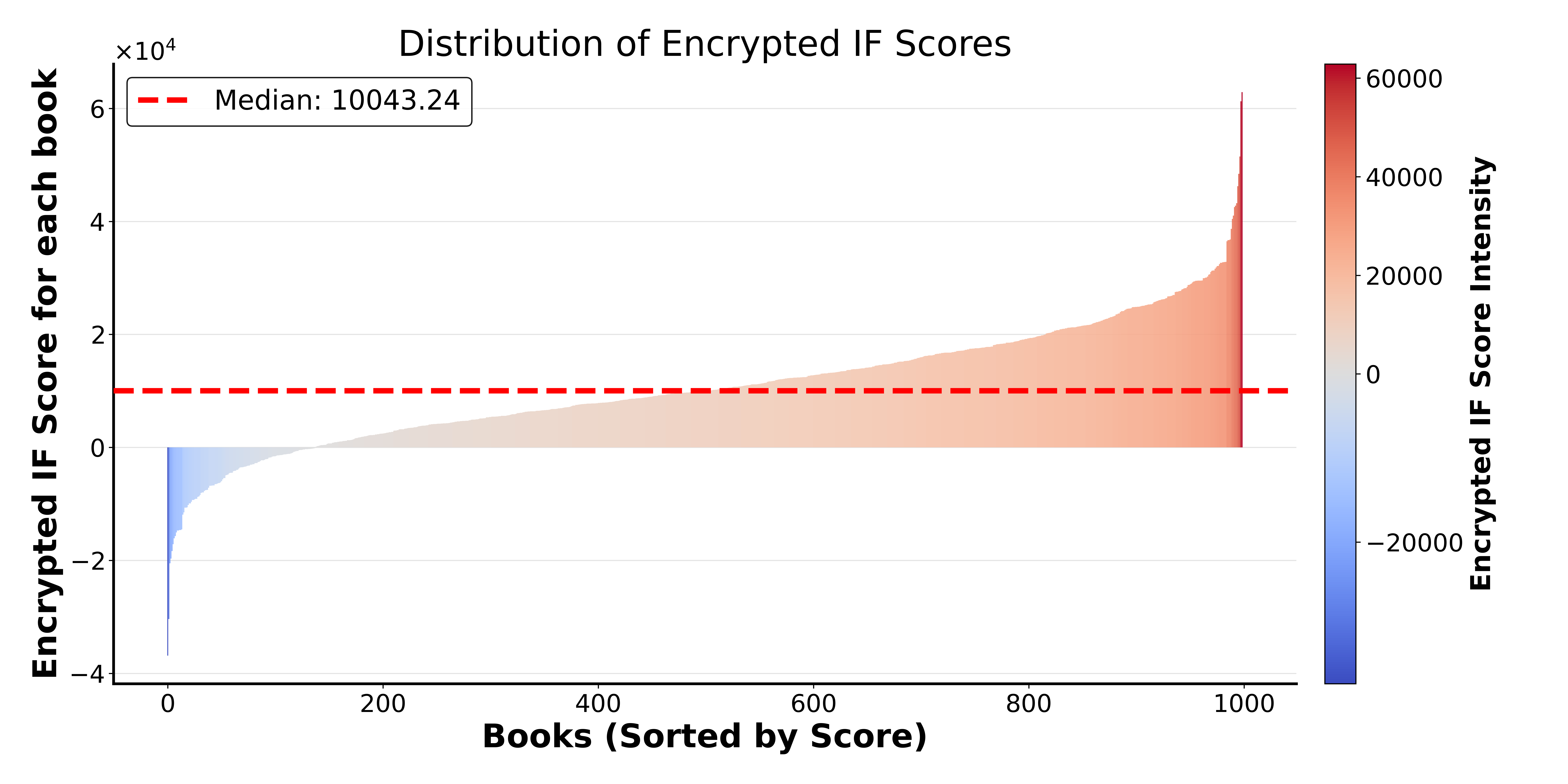} 
    \caption{Distribution of Secure Influence Scores in the Book Marketplace. Each bar represents the encrypted IF score of a book from the dataset. The red dashed line indicates the location of the median of books, sorted by the secure influence score. The distinct skew reveals that a minority of texts drive positive outcomes, while the majority contribute negligible or negative utility, illustrating the inefficiency of uniform pricing.} \label{fig:bookmkt}
\end{figure}

\Cref{fig:bookmkt} presents the ranked distribution of influence scores across the 1,000 sellers. The empirical data reveal a stark asymmetry in data utility that challenges the economic viability of a flat-rate compensation model. The vast majority of the corpus exhibits positive influence scores. In the context of loss minimization, a positive score implies that training on these texts actively increases the buyer's error on the downstream reasoning task. Conversely, utility is concentrated entirely in the minority of the corpus in the rightmost tail. Under a uniform pricing regime like the Bartz settlement, the market forces the buyer to cross-subsidize the ``harmful majority,'' paying for data that degrades model performance at the expense of the sparse minority of authors whose works actually drive capability.

To explain the variance in marginal utility, we analyze representative books. The valuation signal effectively discriminates based on the structural alignment between the training text and the downstream reasoning task. First, the texts identified as most valuable are characterized by explicit entity mapping and syntactic precision, which reinforce the logical scaffolding required for the MuSR team allocation task. 
The highest-utility seller, \textit{A Woman of No Importance (Director's Playbook Edition)} (IF $\approx -36,839$), is a theatrical script. The structure of a play is rigidly mapping character names to dialogue and actions, which is structurally isomorphic to the team allocation task, which requires mapping individuals to roles under constraints. 
By training on this format, the model refines its ability to track entity-state relationships. Similarly, \textit{200 Most Frequently Used French Words} (IF $\approx -30,368$), a text composed of short, literal translation pairs, unexpectedly yields high utility. 
In the context of reasoning, this content likely reinforces concise, unambiguous token-level attention patterns. Unlike narrative prose, which allows for ambiguity, translation pairs demand strict correspondence, steering gradients toward the clarity and precision required for logic puzzles. 
Finally, character-driven narrative works such as \textit{A Lady of Many Charms and Other Stories} (IF $\approx -20,524$) contribute likely due to their dense multi-character interactions, implicit preference signaling, and evolving interpersonal dynamics. These texts require the model to infer latent traits (e.g., competence, trustworthiness, compatibility) from soft contextual clues and dialogues rather than explicit labels. Such narrative structures closely mirror the team allocation task, where successful performance depends on extracting hidden skills and relational constraints from natural language and integrating them into coherent assignment decisions under compatibility constraints.

Conversely, texts that increase the buyer's loss are characterized by semantic drift and ambiguity that compete with the reasoning objective. The most unhelpful seller, \textit{Al-Jaami Al-Saheeh} (IF $\approx +62,866$), is characterized by archaic phrasing and complex theological abstractions. This creates a significant distributional shift away from the distribution of the evaluation set, introducing noise into the model's embeddings. Similarly, \textit{200 Mulla Nasrudin Stories and Jokes} (IF $\approx +61,253$) acts as a functional distractor. The logic of humor often relies on subverting expectations and embracing ambiguity, which is the inverse of the strict constraint satisfaction required for the team allocation task. Interestingly, highly technical content can also be harmful if the domain is misaligned; the \textit{2018 US Combat Aeroplane Accident Compilation} (IF $\approx +51,500$) yields a strong positive influence score. While structured, these dense summaries emphasize mechanical jargon and physics rather than human agency. The hyper-specific vocabulary acts as distracting noise, diluting the model's focus on the interpersonal and managerial reasoning required by the downstream objective.

These findings suggest that secure data valuation can serve as the technical foundation for a fairer data economy. By empowering buyers to identify the specific authors whose works drive model capabilities, the framework shifts the market equilibrium from volume extraction to quality exchange. This enables a licensing model where compensation is proportional to utility, resolving the fairness concerns inherent in current class-action resolutions and the opacity of current data exchange platforms.

\section{Discussion}
\label{sec:discussion}

The experiments above demonstrate that encrypted influence scores can closely track realized utility in settings where controlled retraining is feasible, and can reveal substantial heterogeneity in settings where exhaustive retraining is infeasible. At the same time, the approach rests on modeling choices that warrant explicit clarification for correct interpretation and deployment. First, influence-based valuation is a local, first-order approximation, so its accuracy depends on the acquisition being small relative to the buyer’s existing training set. Second, TIP is described under an implementation in which the model owner can compute the gradients and the preconditioned evaluation direction underlying the influence. The following subsections discuss these boundary conditions and explain how the accessibility requirement can be relaxed in practice. Finally, we outline additional institutional and technical components needed to extend TIP into a full-fledged data marketplace.

\subsection{Boundary Condition}

\subsubsection{Acquisition for Large Datasets}

Our influence-based data evaluation relies on a first-order Taylor expansion around the current model parameters $\hat{\theta}$. While efficient, this approximation implies a strict boundary condition regarding the size of the data acquisition relative to the buyer's existing knowledge base. We formalize this constraint by analyzing the approximation error induced by finite perturbations.

Initially, the buyer minimizes the empirical risk over their existing dataset $D_{train}$ of size $n$: $R_{\text{train}}(\theta) = \frac{1}{n} \sum_{z \in D_{\text{train}}} \ell(z, \theta)$.
When the buyer acquires a candidate set $S$ of size $|S|$, the actual new training objective is the average loss over the combined union of datasets. Let $R_{new}(\theta)$ denote this objective:

\begin{align*}
    R_{new}(\theta) &= \frac{1}{n+|S|} \left[ \sum_{z \in D_{\text{train}}} \ell(z, \theta) + \sum_{z \in S} \ell(z, \theta) \right]\\
    &= \frac{n}{n+|S|} R_{\text{train}}(\theta) + \frac{|S|}{n+|S|} R_S(\theta)
\end{align*}

To map this to the influence framework, we scale the objective by $\frac{n+|S|}{n}$, yielding a proxy objective:
\begin{equation}
    R_{\epsilon}(\theta) = R_{\text{train}}(\theta) + \epsilon R_S(\theta), 
\end{equation}
where the perturbation weight is $\epsilon = \frac{|S|}{n}$. 
Let $\hat{\theta}(\epsilon) = \arg\min_{\theta} R_{\epsilon}(\theta)$ denote the trajectory of the optimal parameters as a function of the perturbation weight $\epsilon$. Note that $\hat{\theta}(0)$ corresponds to the buyer's current parameters $\hat{\theta}$.

The influence function estimates the parameter shift $\Delta \theta$ via a linear projection along the gradient of the loss. Mathematically, this corresponds to the first-order term of the Taylor expansion of the optimal parameters $\hat{\theta}(\epsilon)$ around $\epsilon=0$. However, the true parameter trajectory is curvilinear, governed by the higher-order geometry of the loss landscape. The discrepancy between the linear prediction and the true minimizer is dominated by the second-order Lagrange remainder:
\begin{equation}
    \hat{\theta}(\epsilon) = \hat{\theta}(0) + \epsilon \cdot \frac{d\hat{\theta}}{d\epsilon}\bigg|_{\epsilon=0} + \frac{1}{2}\epsilon^2 \cdot \frac{d^2\hat{\theta}}{d\epsilon^2}\bigg|_{\xi},
\end{equation}
where $\xi \in (0, \epsilon)$ is an intermediate scalar determined by the mean value theorem. The influence score acts as a precise predictor of loss reduction in the incremental acquisition regime ($|S| \ll n$), where the perturbation ratio is sufficiently small that the quadratic error term remains negligible. Conversely, for bulk acquisitions where $|S|$ approaches $n$, the quadratic error dominates, and the first-order approximation may fail to capture the geometric shift of the parameters. Accordingly, for large acquisitions, our influence-based evaluator should be interpreted primarily as a local screening or ranking signal computed at the buyer’s current model state, rather than as a precise predictor of the full retraining effect.

\subsection{Model Accessibility Assumption}
\label{extension:modelaccess}

Our Trustworthy Influence Protocol (TIP) is presented in a white-box setting in which the data buyer (model owner) can compute the quantities required for influence-based evaluation at the buyer’s current trained parameters $\hat{\theta}$. In practice, however, many data buyers operate proprietary models and are unwilling to disclose model parameters to other market agents. TIP is compatible with this constraint because influence-based scoring does not require the model itself to be shared. Instead, it requires that the model owner can compute the influence quantities internally and expose only the minimal outputs needed by the protocol. Concretely, the model owner can keep the model entirely private while computing the relevant gradients and the preconditioned evaluation direction within a trusted execution environment, and then see the resulting influence scores. In this sense, the practical accessibility requirement is gradient access for the model owner, rather than complete open weight access for external parties.

When model access is fully restricted, although coarser, the buyer can still fall back to model-agnostic valuation signals such as Data-Cosine (Section~\ref{sec:casestudy}), which our healthcare simulation shows to provide a meaningful ranking. 
The above requirement can be further relaxed when exact gradients are not directly available to the entity executing the scoring procedure. If the model can be queried to obtain outputs, local sensitivities can be approximated using finite-difference and related zeroth-order gradient estimators based on controlled perturbations. For instance, \citet{nesterov2017random} approximates gradients by applying Gaussian smoothing to the objective and constructing random-direction finite-difference oracles, and \citet{ghadimi2013stochastic} develops stochastic zeroth-order methods that estimate gradients via random-direction finite differences of noisy function evaluations, enabling gradient-based optimization when only value queries are available.
This relaxation increases computational cost because it requires repeated queries and introduces estimation error due to noisy gradients.

Another complementary option under restrictive access is to use a differentiable surrogate model trained by the model owner. The model owner can train a smaller model to mimic the deployed model on the relevant input distribution and then compute influence scores on the surrogate using standard differentiation, allowing external parties to compute gradients or sensitivities with respect to inputs using the surrogate, rather than the inaccessible original model. 
Recent work has explored this idea in various domains.
\citet{Tseng2019Hyper} trained a differentiable proxy model to mimic an arbitrary black-box image signal processing. \citet{khondaker2025learning} used a differentiable surrogate model to approximate the behavior of the image denoising method. 
Such surrogate models, when provided by the model owner, allow researchers to compute gradients or sensitivities as if they had full access, thereby overcoming the restrictions of the original black-box in a safe, controlled manner.

\section{Conclusion}
\label{sec:conclusion}

This paper addresses the value-privacy dilemma in AI data markets: the conundrum of verifying data utility without exposing the data itself. We introduce the Trustworthy Influence Protocol (TIP), a privacy-preserving framework that integrates gradient-based influence functions with Homomorphic Encryption to enable prospective data buyers to quantify the marginal utility of external data while the data remain fully encrypted end-to-end. A key design insight is that the influence-function formula naturally decomposes into a buyer-side component and a seller-side component coupled only through an inner product, a bilinear structure that maps directly onto efficient homomorphic evaluation.

We demonstrate TIP's fidelity and scalability across three models of varying complexity, showing that encrypted influence scores achieve near-perfect correlation with their plaintext counterparts while incurring per-sample overhead that scales with the projected gradient dimension rather than the underlying model size. 
We further validate the framework in two applied settings. In a healthcare data market built from inpatient records, we compare three tiers of valuation signals and show that our method significantly outperforms similarity-based alternatives and effectively resolves adverse selection in such a high-stakes domain. In a generative AI book market, we find that data utility is highly skewed: a minority of texts drive model performance, while the majority degrade it, challenging the prevailing flat-rate compensation schemes.

The framework is subject to several limitations. As we discussed in Section~\ref{sec:discussion}, influence-based valuation relies on a first-order Taylor approximation and is most accurate when the candidate acquisition is small relative to the buyer's existing training set; for large bulk acquisitions, the score should be interpreted as a ranking signal rather than a precise predictor of retraining outcome. Second, TIP requires that the model owner can compute gradients and the preconditioned evaluation vector, though this requirement can be relaxed via zeroth-order gradient estimation or differentiable surrogate models when direct gradient access is unavailable. Third, the CKKS encryption scheme introduces a trade-off between numerical precision and computational cost; the choice of scale parameter directly impacts fidelity, necessitating careful calibration.

Moreover, TIP addresses the core valuation challenge of computing data utility without revealing data, but deploying it within a fully functioning marketplace requires additional components that we have not formally modeled. 
In the current protocol, the broker performs homomorphic computation on encrypted inputs and returns encrypted results, learning nothing about the underlying data, the model, or the scores. The broker's incentive to participate derives not from access to data but from providing computational infrastructure: its compensation can take the form of transaction fees or platform subscriptions, analogous to data intermediaries that profit from brokering information flows rather than from the information itself \citep{bergemann2022economics}. 
Alternatively, the broker role can be assumed by a regulated entity such as a government-chartered data trust or an industry consortium whose mandate is to facilitate secure exchange. Looking further ahead, recent advances in multi-key Fully Homomorphic Encryption \citep{lopez2012multikey, chen2019efficient} suggest that the buyer and seller could jointly compute the encrypted inner product without delegating computation to a third party. While current multi-key HE protocols incur higher communication overhead compared to the single-key setting used in TIP, continued improvements in cryptographic efficiency may render fully peer-to-peer valuation practical, reducing the trust assumptions to their theoretical minimum.

Beyond the broker question, building a full-fledged data marketplace requires complementary institutional mechanisms that translate valuation signals into enforceable economic outcomes. These include incentive-compatible mechanism design that aligns the interests of buyers, sellers, and intermediaries \citep{bergemann2018design, zhang2025fairshare}; payment and settlement protocols that govern how compensation flows from buyer to seller conditional on realized utility, potentially automated via privacy-preserving smart contracts \citep{kosba2016hawk}; and participant identity protection in domains where even the act of offering data may reveal strategic information, addressable through anonymous credential schemes \citep{camenisch2004signature}. While TIP provides the critical cryptographic primitive for the first stage, the design of these surrounding mechanisms constitutes a rich agenda for future research.

\bibliographystyle{unsrtnat}
\bibliography{references} %

@article{zhang2024survey,
  title={A Survey on Data Markets},
  author={Zhang, Jiayao and Bi, Yuran and Cheng, Mengye and Liu, Jinfei and Ren, Kui and Sun, Qiheng and Wu, Yihang and Cao, Yang and Fernandez, Raul Castro and Xu, Haifeng and others},
  journal={arXiv preprint arXiv:2411.07267},
  year={2024}
}

@misc{grynbaum2023times,
  title={The Times sues OpenAI and Microsoft over AI use of copyrighted work},
  author={Grynbaum, Michael M and Mac, Ryan},
  howpublished={The New York Times},
  year={2023},
  month=dec,
  day={27}
}

@inproceedings{ghorbani2019data,
  title={Data shapley: Equitable valuation of data for machine learning},
  author={Ghorbani, Amirata and Zou, James},
  booktitle={International conference on machine learning},
  pages={2242--2251},
  year={2019},
  organization={PMLR}
}

@article{marcolla2022survey,
  title={Survey on fully homomorphic encryption, theory, and applications},
  author={Marcolla, Chiara and Sucasas, Victor and Manzano, Marc and Bassoli, Riccardo and Fitzek, Frank HP and Aaraj, Najwa},
  journal={Proceedings of the IEEE},
  volume={110},
  number={10},
  pages={1572--1609},
  year={2022},
  publisher={IEEE}
}

@article{hu2024most,
  title={Most influential subset selection: Challenges, promises, and beyond},
  author={Hu, Yuzheng and Hu, Pingbang and Zhao, Han and Ma, Jiaqi},
  journal={Advances in Neural Information Processing Systems},
  volume={37},
  pages={119778--119810},
  year={2024}
}

@article{balazinska2011data,
  title={Data markets in the cloud: An opportunity for the database community},
  author={Balazinska, Magdalena and Howe, Bill and Suciu, Dan},
  journal={Proceedings of the VLDB Endowment},
  volume={4},
  number={12},
  pages={1482--1485},
  year={2011},
  publisher={VLDB Endowment}
}

@inproceedings{chen2019towards,
  title={Towards model-based pricing for machine learning in a data marketplace},
  author={Chen, Lingjiao and Koutris, Paraschos and Kumar, Arun},
  booktitle={Proceedings of the 2019 international conference on management of data},
  pages={1535--1552},
  year={2019}
}

@inproceedings{agarwal2019marketplace,
  title={A marketplace for data: An algorithmic solution},
  author={Agarwal, Anish and Dahleh, Munther and Sarkar, Tuhin},
  booktitle={Proceedings of the 2019 ACM Conference on Economics and Computation},
  pages={701--726},
  year={2019}
}

@inproceedings{jia2019towards,
  title={Towards efficient data valuation based on the shapley value},
  author={Jia, Ruoxi and Dao, David and Wang, Boxin and Hubis, Frances Ann and Hynes, Nick and G{\"u}rel, Nezihe Merve and Li, Bo and Zhang, Ce and Song, Dawn and Spanos, Costas J},
  booktitle={The 22nd International Conference on Artificial Intelligence and Statistics},
  pages={1167--1176},
  year={2019},
  organization={PMLR}
}

@article{birkhead2025algorithms,
  title={Algorithms to the Rescue: Market Mechanisms for Consensual Trading of Unbiased Individual Data},
  author={Birkhead, Brian and Eshghi, Ashkan and Gopal, Ram D and Hidaji, Hooman and Patterson, Raymond A},
  journal={Information Systems Research},
  year={2025},
  publisher={INFORMS}
}

@inproceedings{brakerski2012fully,
  title={Fully homomorphic encryption without modulus switching from classical GapSVP},
  author={Brakerski, Zvika},
  booktitle={Annual cryptology conference},
  pages={868--886},
  year={2012},
  organization={Springer}
}

@misc{wang2023security,
  title={Security and privacy on generative data in AIGC: A survey},
  author={Wang, T and Zhang, Y and Qi, S and Zhao, R and Xia, Z and Weng, J},
  year={2023},
  journal={arXiv preprint arXiv:2309.09435},
}

@article{yao2024machine,
  title={Machine unlearning of pre-trained large language models},
  author={Yao, Jin and Chien, Eli and Du, Minxin and Niu, Xinyao and Wang, Tianhao and Cheng, Zezhou and Yue, Xiang},
  journal={arXiv preprint arXiv:2402.15159},
  year={2024}
}

@article{hu2025grass,
  title={GraSS: Scalable Influence Function with Sparse Gradient Compression},
  author={Hu, Pingbang and Melkonian, Joseph and Tang, Weijing and Zhao, Han and Ma, Jiaqi W},
  journal={arXiv preprint arXiv:2505.18976},
  year={2025}
}

@article{li2014theory,
  title={A theory of pricing private data},
  author={Li, Chao and Li, Daniel Yang and Miklau, Gerome and Suciu, Dan},
  journal={ACM Transactions on Database Systems (TODS)},
  volume={39},
  number={4},
  pages={1--28},
  year={2014},
  publisher={Acm New York, NY, USA}
}

@article{deng2025survey,
  title={A Survey of Data Attribution: Methods, Applications, and Evaluation in the Era of Generative AI},
  author={Deng, Junwei and Hu, Yuzheng and Hu, Pingbang and Li, Ting-Wei and Liu, Shixuan and Wang, Jiachen T and Ley, Dan and Dai, Qirun and Huang, Benhao and Huang, Jin and others},
  journal={SSRN Working Paper No.\ 5451054},
  year={2025}
}

@article{koutris2015query,
  title={Query-based data pricing},
  author={Koutris, Paraschos and Upadhyaya, Prasang and Balazinska, Magdalena and Howe, Bill and Suciu, Dan},
  journal={Journal of the ACM (JACM)},
  volume={62},
  number={5},
  pages={1--44},
  year={2015},
  publisher={ACM New York, NY, USA}
}

@ARTICLE{6296535,
  author={Deng, Li},
  journal={IEEE Signal Processing Magazine}, 
  title={The MNIST Database of Handwritten Digit Images for Machine Learning Research [Best of the Web]}, 
  year={2012},
  volume={29},
  number={6},
  pages={141-142},
  doi={10.1109/MSP.2012.2211477}}

@inproceedings{socher-etal-2013-recursive,
    title = "Recursive Deep Models for Semantic Compositionality Over a Sentiment Treebank",
    author = "Socher, Richard  and
      Perelygin, Alex  and
      Wu, Jean  and
      Chuang, Jason  and
      Manning, Christopher D.  and
      Ng, Andrew  and
      Potts, Christopher",
    booktitle = "Proceedings of the 2013 Conference on Empirical Methods in Natural Language Processing",
    month = oct,
    year = "2013",
    address = "Seattle, Washington, USA",
    publisher = "Association for Computational Linguistics",
    url = "https://www.aclweb.org/anthology/D13-1170",
    pages = "1631--1642",
}

@misc{merity2016pointer,
      title={Pointer Sentinel Mixture Models},
      author={Stephen Merity and Caiming Xiong and James Bradbury and Richard Socher},
      year={2016},
      eprint={1609.07843},
      archivePrefix={arXiv},
      primaryClass={cs.CL}
}

@inproceedings{wang2018glue,
  title={GLUE: A multi-task benchmark and analysis platform for natural language understanding},
  author={Wang, Alex and Singh, Amanpreet and Michael, Julian and Hill, Felix and Levy, Omer and Bowman, Samuel},
  booktitle={Proceedings of the 2018 EMNLP workshop BlackboxNLP: Analyzing and interpreting neural networks for NLP},
  pages={353--355},
  year={2018}
}

@article{radford2019language,
  title={Language models are unsupervised multitask learners},
  author={Radford, Alec and Wu, Jeffrey and Child, Rewon and Luan, David and Amodei, Dario and Sutskever, Ilya and others},
  journal={OpenAI blog},
  volume={1},
  number={8},
  pages={9},
  year={2019}
}

@inproceedings{cheon2017homomorphic,
  title={Homomorphic encryption for arithmetic of approximate numbers},
  author={Cheon, Jung Hee and Kim, Andrey and Kim, Miran and Song, Yongsoo},
  booktitle={Advances in cryptology--ASIACRYPT 2017: 23rd international conference on the theory and applications of cryptology and information security, Hong kong, China, December 3-7, 2017, proceedings, part i 23},
  pages={409--437},
  year={2017},
  organization={Springer}
}

@article{roughgarden2010algorithmic,
  title={Algorithmic game theory},
  author={Roughgarden, Tim},
  journal={Communications of the ACM},
  volume={53},
  number={7},
  pages={78--86},
  year={2010},
  publisher={ACM New York, NY, USA}
}

@article{wang2021blockchain,
  title={Blockchain-enabled data sharing in supply chains: Model, operationalization, and tutorial},
  author={Wang, Zhiyuan and Zheng, Zhiqiang and Jiang, Wei and Tang, Shaojie},
  journal={Production and Operations Management},
  volume={30},
  number={7},
  pages={1965--1985},
  year={2021},
  publisher={SAGE Publications Sage CA: Los Angeles, CA}
}

@article{choe2024your,
  title={What is your data worth to gpt? llm-scale data valuation with influence functions},
  author={Choe, Sang Keun and Ahn, Hwijeen and Bae, Juhan and Zhao, Kewen and Kang, Minsoo and Chung, Youngseog and Pratapa, Adithya and Neiswanger, Willie and Strubell, Emma and Mitamura, Teruko and others},
  journal={arXiv preprint arXiv:2405.13954},
  year={2024}
}

@inproceedings{ribeiro2016should,
  title={" Why should i trust you?" Explaining the predictions of any classifier},
  author={Ribeiro, Marco Tulio and Singh, Sameer and Guestrin, Carlos},
  booktitle={ACM SIGKDD},
  pages={1135--1144},
  year={2016}
}

@incollection{NIPS2017_7062,
title = {A Unified Approach to Interpreting Model Predictions},
author = {Lundberg, Scott M and Lee, Su-In},
booktitle = {Advances in Neural Information Processing Systems 30},
editor = {I. Guyon and U. V. Luxburg and S. Bengio and H. Wallach and R. Fergus and S. Vishwanathan and R. Garnett},
pages = {4765--4774},
year = {2017},
publisher = {Curran Associates, Inc.},
url = {http://papers.nips.cc/paper/7062-a-unified-approach-to-interpreting-model-predictions.pdf}
}

@inproceedings{koh2017understanding,
  title={Understanding black-box predictions via influence functions},
  author={Koh, Pang Wei and Liang, Percy},
  booktitle={International conference on machine learning},
  pages={1885--1894},
  year={2017},
  organization={PMLR}
}

@article{zhang2020survey,
  title={A survey of data pricing methods},
  author={Zhang, Mengxiao and Beltr{\'a}n, Fernando},
  journal={SSRN J},
  year={2020}
}

@article{fernandez2020data,
  title={Data market platforms: Trading data assets to solve data problems},
  author={Fernandez, Raul Castro and Subramaniam, Pranav and Franklin, Michael J},
  journal={arXiv preprint arXiv:2002.01047},
  year={2020}
}

@article{fan2012somewhat,
  title={Somewhat practical fully homomorphic encryption},
  author={Fan, Junfeng and Vercauteren, Frederik},
  journal={Cryptology ePrint Archive},
  year={2012}
}

@inproceedings{park2023trak,
  title={TRAK: Attributing Model Behavior at Scale},
  author={Park, Sung Min and Georgiev, Kristian and Ilyas, Andrew and Leclerc, Guillaume and Madry, Aleksander},
  booktitle={arXiv preprint arXiv:2303.14186},
  year={2023}
}

@book{dwork2014algorithmic,
  author    = {Cynthia Dwork and Aaron Roth},
  title     = {The Algorithmic Foundations of Differential Privacy},
  publisher = {Foundations and Trends® in Theoretical Computer Science},
  year      = {2014}
}

@article{spiekermann2019data,
  title={Data marketplaces: Trends and monetisation of data goods},
  author={Spiekermann, Markus},
  journal={Intereconomics},
  volume={54},
  number={4},
  pages={208--216},
  year={2019},
  publisher={Springer}
}

@misc{Brittain2023_Getty_Reuters,
  author       = {Brittain, Blake},
  title        = {Getty Images lawsuit says Stability AI misused photos to train {AI}},
  howpublished      = {Reuters},
  year         = {2023},
  month        = feb,
  day          = {6},
  url          = {https://www.reuters.com/legal/getty-images-lawsuit-says-stability-ai-misused-photos-train-ai-2023-02-06/}
}

@misc{OpenAI2024_RedditPartnership,
  author       = {{OpenAI} and {Reddit}},
  title        = {OpenAI and Reddit Partnership},
  howpublished = {OpenAI Blog},
  year         = {2024},
  month        = may,
  day          = {16},
  url          = {https://openai.com/index/openai-and-reddit-partnership/}
}

@inproceedings{ghosh2011selling,
  title={Selling privacy at auction},
  author={Ghosh, Arpita and Roth, Aaron},
  booktitle={Proceedings of the 12th ACM conference on Electronic commerce},
  pages={199--208},
  year={2011}
}

@article{sprague2023musr,
  title={Musr: Testing the limits of chain-of-thought with multistep soft reasoning},
  author={Sprague, Zayne and Ye, Xi and Bostrom, Kaj and Chaudhuri, Swarat and Durrett, Greg},
  journal={arXiv preprint arXiv:2310.16049},
  year={2023}
}

@article{broderick2020automatic,
  title={An automatic finite-sample robustness metric: when can dropping a little data make a big difference?},
  author={Broderick, Tamara and Giordano, Ryan and Meager, Rachael},
  journal={arXiv preprint arXiv:2011.14999},
  year={2020}
}

@article{koh2019accuracy,
  title={On the accuracy of influence functions for measuring group effects},
  author={Koh, Pang Wei W and Ang, Kai-Siang and Teo, Hubert and Liang, Percy S},
  journal={Advances in neural information processing systems},
  volume={32},
  year={2019}
}

@incollection{arrow1962economic,
  title={Economic welfare and the allocation of resources for invention},
  author={Arrow, Kenneth Joseph},
  booktitle={Readings in industrial economics: Volume two: Private enterprise and state intervention},
  pages={219--236},
  year={1962},
  publisher={Springer}
}

@article{SmithDinevXu2011,
  author  = {Smith, H. J. and Dinev, T. and Xu, H.},
  title   = {Information privacy research: an interdisciplinary review},
  journal = {MIS Quarterly},
  pages   = {989--1015},
  year    = {2011}
}

@article{LiSarkar2011,
  author  = {Li, X. B. and Sarkar, S.},
  title   = {Protecting privacy against record linkage disclosure: A bounded swapping approach for numeric data},
  journal = {Information Systems Research},
  volume  = {22},
  number  = {4},
  pages   = {774--789},
  year    = {2011}
}

@article{LiQin2017,
  author  = {Li, X. B. and Qin, J.},
  title   = {Anonymizing and sharing medical text records},
  journal = {Information Systems Research},
  volume  = {28},
  number  = {2},
  pages   = {332--352},
  year    = {2017}
}

@article{XuDinev2022,
  author  = {Xu, H. and Dinev, T.},
  title   = {Guest editorial: Reflections on the 2021 impact award: Why privacy still matters},
  journal = {MIS Quarterly},
  volume  = {46},
  number  = {4},
  pages   = {xx--xxxii},
  year    = {2022}
}

@article{XuTeoTanAgarwal2012,
  author  = {Xu, H. and Teo, H. H. and Tan, B. C. and Agarwal, R.},
  title   = {Research note---effects of individual self-protection, industry self-regulation, and government regulation on privacy concerns: a study of location-based services},
  journal = {Information Systems Research},
  volume  = {23},
  number  = {4},
  pages   = {1342--1363},
  year    = {2012}
}

@article{BuckmanBockstedtHashim2019,
  author  = {Buckman, J. R. and Bockstedt, J. C. and Hashim, M. J.},
  title   = {Relative privacy valuations under varying disclosure characteristics},
  journal = {Information Systems Research},
  volume  = {30},
  number  = {2},
  pages   = {375--388},
  year    = {2019}
}

@article{XuZhang2022,
  author  = {Xu, H. and Zhang, N.},
  title   = {From contextualizing to context theorizing: Assessing context effects in privacy research},
  journal = {Management Science},
  volume  = {68},
  number  = {10},
  pages   = {7383--7401},
  year    = {2022}
}

@article{yang2025secure,
  title={Secure Confidential Business Information When Sharing Machine Learning Models},
  author={Yang, Yunfan and Xu, Jiarong and Zhang, Hongzhe and Fang, Xiao},
  journal={arXiv preprint arXiv:2509.16352},
  year={2025}
}

@article{HanWangWuFang2025,
  author  = {Han, X. and Wang, L. and Wu, J. and Fang, X.},
  title   = {Data valuation for vertical federated learning: A model-free and privacy-preserving method},
  journal = {MIS Quarterly},
  note    = {forthcoming},
  year    = {2025}
}

@article{grattafiori2024llama,
  title={The llama 3 herd of models},
  author={Grattafiori, Aaron and Dubey, Abhimanyu and Jauhri, Abhinav and Pandey, Abhinav and Kadian, Abhishek and Al-Dahle, Ahmad and Letman, Aiesha and Mathur, Akhil and Schelten, Alan and Vaughan, Alex and others},
  journal={arXiv preprint arXiv:2407.21783},
  year={2024}
}

@article{zhang2025fairshare,
  title={Fairshare Data Pricing via Data Valuation for Large Language Models},
  author={Zhang, Luyang and Jiao, Cathy and Li, Beibei and Xiong, Chenyan},
  journal={arXiv preprint arXiv:2502.00198},
  year={2025}
}

@article{liu2024deepseek,
  title={Deepseek-v3 technical report},
  author={Liu, Aixin and Feng, Bei and Xue, Bing and Wang, Bingxuan and Wu, Bochao and Lu, Chengda and Zhao, Chenggang and Deng, Chengqi and Zhang, Chenyu and Ruan, Chong and others},
  journal={arXiv preprint arXiv:2412.19437},
  year={2024}
}

@article{cobbe2021gsm8k,
  title={Training Verifiers to Solve Math Word Problems},
  author={Cobbe, Karl and Kosaraju, Vineet and Bavarian, Mohammad and Chen, Mark and Jun, Heewoo and Kaiser, Lukasz and Plappert, Matthias and Tworek, Jerry and Hilton, Jacob and Nakano, Reiichiro and Hesse, Christopher and Schulman, John},
  journal={arXiv preprint arXiv:2110.14168},
  year={2021}
}

@misc{wang2024mmlu,
  title         = {MMLU-Pro: A More Robust and Challenging Multi-Task Language Understanding Benchmark},
  author        = {Wang, Yubo and others},
  year          = {2024},
  eprint        = {2406.01574},
  archivePrefix = {arXiv},
  primaryClass  = {cs.CL},
  url           = {https://arxiv.org/abs/2406.01574}
}

@article{pearlmutter1994fast,
  title={Fast exact multiplication by the Hessian},
  author={Pearlmutter, Barak A},
  journal={Neural computation},
  volume={6},
  number={1},
  pages={147--160},
  year={1994},
  publisher={MIT Press}
}

@inproceedings{martens2010deep,
  title={Deep learning via hessian-free optimization.},
  author={Martens, James and others},
  booktitle={Icml},
  volume={27},
  pages={735--742},
  year={2010}
}

@article{agarwal2017second,
  title={Second-order stochastic optimization for machine learning in linear time},
  author={Agarwal, Naman and Bullins, Brian and Hazan, Elad},
  journal={Journal of Machine Learning Research},
  volume={18},
  number={116},
  pages={1--40},
  year={2017}
}

@article{Tseng2019Hyper,
title = {Hyperparameter Optimization in Black-box Image Processing using Differentiable Proxies},
author = {Tseng, Ethan and Yu, Felix and Yang, Yuting and Mannan, Fahim and Arnaud, Karl St. and Nowrouzezahrai, Derek and Lalonde, Jean-Francois and Heide, Felix},
journal = {ACM Transactions on Graphics (TOG)},
doi = {10.1145/3306346.3322996},
volume = {38},
number = {4},
year = {2019},
month = {7},
publisher = {ACM}
}

@inproceedings{khondaker2025learning,
  title={Learning Instance-Specific Parameters of Black-Box Models Using Differentiable Surrogates},
  author={Khondaker, Arnisha and Ray, Nilanjan},
  booktitle={2025 IEEE/CVF Winter Conference on Applications of Computer Vision (WACV)},
  pages={7429--7438},
  year={2025},
  organization={IEEE}
}

@article{nesterov2017random,
  title={Random gradient-free minimization of convex functions},
  author={Nesterov, Yurii and Spokoiny, Vladimir},
  journal={Foundations of Computational Mathematics},
  volume={17},
  number={2},
  pages={527--566},
  year={2017},
  publisher={Springer}
}

@article{ghadimi2013stochastic,
  title={Stochastic first-and zeroth-order methods for nonconvex stochastic programming},
  author={Ghadimi, Saeed and Lan, Guanghui},
  journal={SIAM journal on optimization},
  volume={23},
  number={4},
  pages={2341--2368},
  year={2013},
  publisher={SIAM}
}

@article{lindell2020secure,
  title={Secure multiparty computation},
  author={Lindell, Yehuda},
  journal={Communications of the ACM},
  volume={64},
  number={1},
  pages={86--96},
  year={2020},
  publisher={ACM New York, NY, USA}
}

@techreport{HomomorphicEncryptionSecurityStandard,
 author = {Martin Albrecht and Melissa Chase and Hao Chen and Jintai Ding and Shafi Goldwasser and Sergey Gorbunov and Shai Halevi and Jeffrey Hoffstein and Kim Laine and Kristin Lauter and Satya Lokam and Daniele Micciancio and Dustin Moody and Travis Morrison and Amit Sahai and Vinod Vaikuntanathan},
 title = {Homomorphic Encryption Security Standard},
 institution= {HomomorphicEncryption.org},
 publisher = {HomomorphicEncryption.org},
 address = {Toronto, Canada},
 year = {2018},
 month = {November}
 }

@article{ghoshal2020hiding,
  title={Hiding sensitive information when sharing distributed transactional data},
  author={Ghoshal, Abhijeet and Hao, Jing and Menon, Syam and Sarkar, Sumit},
  journal={Information systems research},
  volume={31},
  number={2},
  pages={473--490},
  year={2020},
  publisher={INFORMS}
}

@article{menon2016privacy,
  title={Privacy and Big Data: Scalable Approaches to Sanitize Large Transactional Databases for Sharing1},
  author={Menon, Syam and Sarkar, Sumit},
  journal={MIS Quarterly},
  volume={40},
  number={4},
  pages={963--981},
  year={2016},
  publisher={Management Information Systems Research Center, University of Minnesota}
}

@misc{meta2024llama3,
    title = {Llama 3 Model Card},
    author = {{Meta AI}},
    year = {2024},
    howpublished = {GitHub},
    url = {https://github.com/meta-llama/llama3/blob/main/MODEL_CARD.md},
    note = {Accessed: 2026-02-22}
}

@misc{openai2024finetuning,
    title = {Create a fine-tuned model},
    author = {{OpenAI}},
    year = {2024},
    howpublished = {OpenAI Platform Documentation},
    url = {https://platform.openai.com/docs/guides/fine-tuning/create-a-fine-tuned-model},
    note = {Accessed: 2026-02-22}
}

@misc{aws2024dataexchange,
    title = {{AWS Data Exchange}},
    author = {{Amazon Web Services}},
    year = {2024},
    howpublished = {Web page},
    url = {https://aws.amazon.com/data-exchange/}
}

@misc{bartz2024anthropic,
    title = {Bartz v. Anthropic PBC, Case No. 3:24-cv-04760},
    author = {{U.S. District Court for the Northern District of California}},
    year = {2024},
    howpublished = {CourtListener},
    url = {https://www.courtlistener.com/docket/69058235/bartz-v-anthropic-pbc/}
}

@misc{huggingface2024bookcorpus,
    title = {{BookCorpusOpen} Dataset},
    author = {Diliello, Luca},
    year = {2024},
    howpublished = {Hugging Face},
    url = {https://huggingface.co/datasets/lucadiliello/bookcorpusopen}
}

@misc{HangzhouInternetCourt2024_Ultraman_GenAIPlatform,
  author       = {{Hangzhou Internet Court}},
  title        = {Shanghai Cultural Development Co., Ltd. v. Hangzhou Intelligent Technology Co., Ltd., Civil Judgment (2024) Zhe 0192 Min Chu No. 1587 (Generative AI Platform Copyright Case)},
  howpublished = {Civil judgment, Hangzhou Internet Court},
  year         = {2024},
  month        = sep,
  day          = {25},
  url          = {https://www.ciplawyer.cn/articles/155988.html},
  note         = {Accessed 2026-02-22}
}

@misc{Bibas2025_ThomsonReuters_v_Ross_MemorandumOpinion,
  author       = {Bibas, Stephanos},
  title        = {Memorandum Opinion, Thomson Reuters Enterprise Centre GmbH et al. v. Ross Intelligence Inc., No. 1:20-cv-613-SB (D. Del.)},
  howpublished = {United States District Court for the District of Delaware},
  year         = {2025},
  month        = feb,
  day          = {11},
  url          = {https://www.ded.uscourts.gov/sites/ded/files/opinions/20-613_5.pdf},
  note         = {Partial summary judgment on copyright infringement and fair use; Accessed 2026-02-22}
}

@misc{LGMunichI2025_GEMA_OpenAI_Memorisation_Judgment,
  author       = {{Landgericht M{\"u}nchen I}},
  title        = {Unzul{\"a}ssige Vervielf{\"a}ltigung durch Memorisierung von Werken im und durch KI-Sprachmodell (Endurteil v. 11.11.2025 -- 42 O 14139/24)},
  howpublished = {Judgment (Endurteil), Bayern.Recht / gesetze-bayern.de},
  year         = {2025},
  month        = nov,
  day          = {11},
  url          = {https://www.gesetze-bayern.de/Content/Document/Y-300-Z-GRURRS-B-2025-N-30204},
  note         = {Accessed 2026-02-22}
}

@misc{reuters2025meta,
  author       = {{Reuters}},
  title        = {Meta strikes multiple {AI} deals with news publishers, {Axios} reports},
  howpublished = {Reuters},
  year         = {2025},
  month        = dec,
  day          = {5},
  url          = {https://www.reuters.com/business/meta-strikes-multiple-ai-deals-with-news-publishers-axios-reports-2025-12-05/},
}

@misc{AP2023_OpenAI_APArchive,
  author       = {{The Associated Press}},
  title        = {AP, OpenAI Agree to Share Select News Content and Technology in New Collaboration},
  howpublished = {AP Press Release},
  year         = {2023},
  month        = jul,
  day          = {13},
  url          = {https://www.ap.org/media-center/press-releases/2023/ap-open-ai-agree-to-share-select-news-content-and-technology-in-new-collaboration/},
}

@misc{Reuters2023_OpenAI_AxelSpringer,
  author       = {{Reuters}},
  title        = {Global news publisher Axel Springer partners with OpenAI in landmark deal},
  howpublished = {Reuters},
  year         = {2023},
  month        = dec,
  day          = {13},
  url          = {https://www.reuters.com/business/media-telecom/global-news-publisher-axel-springer-partners-with-openai-landmark-deal-2023-12-13/},
}

@misc{OpenAI2024_TIME_Partnership,
  author       = {{OpenAI}},
  title        = {Strategic Content Partnership with TIME},
  howpublished = {OpenAI Newsroom Post},
  year         = {2024},
  month        = jun,
  day          = {27},
  url          = {https://openai.com/index/strategic-content-partnership-with-time/},
}

@misc{NewsCorp2024_OpenAI_Partnership,
  author       = {{News Corp}},
  title        = {News Corp and OpenAI Sign Landmark Multi-Year Global Partnership},
  howpublished = {Investor Relations Press Release},
  year         = {2024},
  month        = may,
  day          = {22},
  url          = {https://investors.newscorp.com/news-releases/news-release-details/news-corp-and-openai-sign-landmark-multi-year-global-partnership},
}

@misc{AnthropicSettlement_FAQ,
  author       = {{Anthropic Copyright Settlement}},
  title        = {Frequently Asked Questions (FAQ)},
  howpublished = {Settlement administration website for Bartz et al. v. Anthropic PBC},
  year         = {2026},
  url          = {https://www.anthropiccopyrightsettlement.com/faq},
  note         = {Accessed 2026-02-22}
}

@misc{AnthropicSettlement_Dates,
  author       = {{Anthropic Copyright Settlement}},
  title        = {Key Dates},
  howpublished = {Settlement administration website for Bartz et al. v. Anthropic PBC},
  year         = {2026},
  url          = {https://www.anthropiccopyrightsettlement.com/dates},
  note         = {Accessed 2026-02-22}
}

@misc{Google2026_Gemini31Pro,
  author       = {{Google}},
  title        = {Gemini 3.1 Pro},
  howpublished = {Google Blog},
  url          = {https://blog.google/innovation-and-ai/models-and-research/gemini-models/gemini-3-1-pro/},
  note         = {Accessed February 23, 2026},
  year         = {2026},
}

@misc{Anthropic2026_ClaudeSonnet46,
  author       = {{Anthropic}},
  title        = {Claude Sonnet 4.6},
  howpublished = {Anthropic News},
  url          = {https://www.anthropic.com/news/claude-sonnet-4-6},
  note         = {Accessed February 23, 2026},
  year         = {2026},
}

@misc{reuters_bytedance_ai_video_2026,
  author       = {Baptista, Eduardo},
  title        = {ByteDance's new AI video model goes viral as China looks for second DeepSeek moment},
  howpublished = {Reuters},
  year         = {2026},
  month        = feb,
  day          = {12},
  url          = {https://www.reuters.com/business/media-telecom/bytedances-new-ai-video-model-goes-viral-china-looks-second-deepseek-moment-2026-02-12},
  note         = {Accessed 2026-03-01}
}

@misc{thewrap_disney_bytedance_seedance_2026,
  author       = {Bryant, Jacob},
  title        = {Disney Hits ByteDance With Cease-and-Desist Over Seedance 2.0},
  howpublished = {TheWrap},
  year         = {2026},
  month        = feb,
  url          = {https://www.thewrap.com/industry-news/public-policy-legal/disney-bytedance-seedance-cease-and-desist/},
  note         = {Accessed 2026-03-01}
}

@misc{mpa_seedance_cease_2026,
  author       = {Motion Picture Association},
  title        = {Motion Picture Association Calls for ByteDance to Cease Seedance 2.0 Infringing Activity},
  howpublished = {Press release},
  year         = {2026},
  url          = {https://www.motionpictures.org/press/motion-picture-association-calls-for-bytedance-to-cease-seedance-2-0-infringing-activity/},
  note         = {Accessed 2026-03-01}
}

@misc{Salman2024_OpenAI_Deleted_TrainingData,
  author       = {Salman, Ali},
  title        = {OpenAI Accidentally Deletes ChatGPT Training Data Amid Publisher Copyright Claims},
  year         = {2024},
  howpublished = {WccFtech},
  url          = {https://wccftech.com/openai-deleted-chatgpt-training-data/},
  note         = {November 22, 2024}
}

@article{bergemann2022economics,
    author  = {Dirk Bergemann and Alessandro Bonatti and Tan Gan},
    title   = {The Economics of Social Data},
    journal = {RAND Journal of Economics},
    volume  = {53},
    number  = {2},
    pages   = {263--296},
    year    = {2022}
  }

@inproceedings{lopez2012multikey,
    author    = {Adriana L\'{o}pez-Alt and Eran Tromer
                 and Vinod Vaikuntanathan},
    title     = {On-the-Fly Multiparty Computation on the Cloud via
                 Multikey Fully Homomorphic Encryption},
    booktitle = {Proceedings of the 44th Annual ACM Symposium on
                 Theory of Computing (STOC)},
    pages     = {1219--1234},
    year      = {2012}
  }

@inproceedings{chen2019efficient,
    author    = {Hao Chen and Wei Dai and Miran Kim and Yongsoo Song},
    title     = {Efficient Multi-Key Homomorphic Encryption with Packed
                 Ciphertexts with Application to Oblivious Neural
                 Network Inference},
    booktitle = {Proceedings of the 2019 ACM CCS},
    pages     = {395--412},
    year      = {2019}
  }

@article{bergemann2018design,
    author  = {Dirk Bergemann and Alessandro Bonatti and Alex Smolin},
    title   = {The Design and Price of Information},
    journal = {American Economic Review},
    volume  = {108},
    number  = {1},
    pages   = {1--48},
    year    = {2018}
  }

@inproceedings{kosba2016hawk,
    author    = {Ahmed Kosba and Andrew Miller and Elaine Shi
                 and Zikai Wen and Charalampos Papamanthou},
    title     = {Hawk: The Blockchain Model of Cryptography and
                 Privacy-Preserving Smart Contracts},
    booktitle = {2016 IEEE Symposium on Security and Privacy (SP)},
    pages     = {839--858},
    year      = {2016}
  }

@inproceedings{camenisch2004signature,
    author    = {Jan Camenisch and Anna Lysyanskaya},
    title     = {Signature Schemes and Anonymous Credentials from
                 Bilinear Maps},
    booktitle = {Advances in Cryptology -- CRYPTO 2004},
    series    = {LNCS},
    volume    = {3152},
    pages     = {56--72},
    year      = {2004}
  }

\section{Appendix}

\subsection{Lemma~\ref{lem:if-loss}}
\label{app:lemma3-1}

We follow the standard derivation from \citet{koh2017understanding}. The empirical risk for the original model is denoted by $R(\theta)$, with minimizer of $\hat{\theta}$. Then, the perturbed risk by incorporating $z_s$ will be:

\begin{equation}
\label{app:perturbedrisk}
    R_{\epsilon, z_s} = R_{\text{train}}(\theta) + \epsilon \ell(z_s; {\theta}).
\end{equation}

By definition, the perturbed minimizer \(\hat{\theta}_{\epsilon}\) satisfies the first-order optimality condition:

$$
\nabla_\theta R_{\epsilon,z_s}(\hat{\theta}_{\epsilon}) = \nabla_\theta R_{\text{train}}(\hat{\theta}_{\epsilon}) + \epsilon \nabla_\theta \ell(z_s; \hat{\theta}_{\epsilon}) = 0.
$$

Here, we cannot solve the equation above directly because $\hat{\theta}_{\epsilon}$ is unknown. However, since $\epsilon \to 0$, $\hat{\theta}_{\epsilon} \approx \hat{\theta}$. We use Taylor Expansion to linearize the gradients around the original parameters $\hat{\theta}$. First, we perform a first-order Taylor expansion of $\nabla_\theta R$ around $\hat{\theta}$:

$$
\nabla_\theta R(\hat{\theta}_{\epsilon, z_s}) \approx \nabla_\theta R(\hat{\theta}) + \nabla_\theta^2 R(\hat{\theta})\cdot(\hat{\theta}_{\epsilon} - \hat{\theta}) + O(||\hat{\theta}_{\epsilon} - \hat{\theta}||^2).
$$

Since \(\hat{\theta}\) is the local minimizer of the training risk, \(\nabla_\theta R_{\text{train}}(\hat{\theta}) = 0\).The term $\nabla_\theta^2 R_{\text{train}}(\hat{\theta})$ corresponds to the Hessian matrix $H_{\hat{\theta}}$. Substituting this back into the optimality condition and ignoring higher-order terms:
\[
H_{\hat{\theta}}(\hat{\theta}_{\epsilon} - \hat{\theta}) + \epsilon \nabla_\theta \ell(z_s; \hat{\theta}) \approx 0.
\]
Solving for the parameter change \(\Delta \theta = \hat{\theta}_{\epsilon} - \hat{\theta}\):

\[
\Delta \theta \approx -\epsilon H_{\hat{\theta}}^{-1} \nabla_\theta \ell(z_s; \hat{\theta}).
\]

Differentiating with respect to \(\epsilon\) at \(\epsilon=0\) gives the rate of parameter change: \(\frac{d\hat{\theta}}{d\epsilon} = -H_{\hat{\theta}}^{-1} \nabla_\theta \ell(z_s; \hat{\theta})\). Finally, to find the influence on the evaluation point, we apply the chain rule to the evaluation loss \(\ell(z_{\mathrm{eval}}; \theta)\):
\[
\frac{d \ell(z_{\mathrm{eval}}; \hat{\theta}_\epsilon)}{d\epsilon} = \nabla_\theta \ell(z_{\mathrm{eval}}; \hat{\theta})^\top \frac{d\hat{\theta}}{d\epsilon} = -\nabla_\theta \ell(z_{\mathrm{eval}}; \hat{\theta})^\top H_{\hat{\theta}}^{-1} \nabla_\theta \ell(z_s; \hat{\theta}).
\]
This yields the result in Eq.~\eqref{eq:if_loss}.

\subsection{Encryption Mechanism and CKKS Technical Details}
\label{app:ckks}

In this section, we provide the cryptographic definitions and algorithmic details of the CKKS scheme, and the technical background for Lemma~\ref{lem:ckks-approx}. Unlike exact Homomorphic Encryption schemes (e.g., BFV, BGV), CKKS \citep{cheon2017homomorphic} is designed for approximate arithmetic over complex or real numbers, making it especially suitable for statistical analysis and machine learning tasks where small precision losses are acceptable.

The essence of the CKKS encryption algorithm is the Ring Learning With Errors (RLWE) problem, whose security underlies the security of our protocol and the resulting indistinguishability properties of the encryption scheme. The RLWE problem posits that for a secret polynomial $s(X)$ and a small error term $e(X)$ sampled from specific distributions (typically discrete Gaussian), samples of the form $(a, b = a \cdot s + e)$ over the ring are computationally indistinguishable from uniformly random pairs $(a, u)$. Consequently, recovering the secret $s$ from the public key or ciphertexts is mathematically infeasible, provided the polynomial degree $N$ is sufficiently large. Based on this hardness assumption, the CKKS scheme achieves \textit{Indistinguishability under Chosen-Plaintext Attack} (IND-CPA). This property guarantees that a ciphertext reveals no information about the underlying plaintext. Formally, even if an adversary can choose two messages and see the encryption of one, it cannot distinguish which message was encrypted with a probability significantly better than random guessing. In the context of our protocol, IND-CPA ensures that the encrypted gradients and evaluation vectors appear to the Broker and any unauthorized party as unstructured random noise, thereby maintaining the strict confidentiality of data and tasks during computation.

Thus, we first define the polynomial ring $\mathcal{R} = \mathbb{Z}[X]/(X^N + 1)$, where $N$ is a power of two (e.g., $N=2^{14}$). Elements in this ring are polynomials of degree up to $N-1$ with integer coefficients. To support leveled homomorphic operations (i.e., a finite depth of multiplications), define a chain of moduli $q_L > q_{L-1} > \dots > q_0$. A ciphertext at level $l$ is an element of the ring $\mathcal{R}_{q_l} = \mathcal{R}/q_l\mathcal{R}$. The reduction of modulus from $q_l$ to $q_{l-1}$ is central to the rescaling mechanism for noise management.

The fundamental technical challenge in the CKKS scheme is to map a data vector of complex or real numbers $z \in \mathbb{C}^{N/2}$ into the integer polynomial ring $\mathcal{R} = \mathbb{Z}[X]/(X^N + 1)$. This mapping is achieved through the mechanism of canonical embedding, denoted as $\sigma$, which establishes an isomorphic relationship between the polynomial ring and the complex vector space. Mathematically, for a polynomial $a(X) \in \mathcal{R}$, the canonical embedding $\sigma(a)$ is defined as the vector of evaluations of the polynomial at the primitive roots of unity, i.e., $\sigma(a) = (a(\zeta^j))_{j \in T}$, where $\zeta$ is a primitive $M$-th root of unity and $T$ is a specific index set of size $N/2$. This algebraic structure is critical because it enables the Single Instruction, Multiple Data (SIMD) property: arithmetic operations (addition and multiplication) performed on the polynomials in the ring $\mathcal{R}$ correspond directly to component-wise operations on the underlying message vectors in $\mathbb{C}^{N/2}$. Consequently, a single homomorphic operation on a ciphertext simultaneously processes $N/2$ independent data slots, significantly enhancing computational efficiency.

Since the underlying RLWE encryption scheme requires plaintext to be integer polynomials, floating-point data cannot be directly embedded without modification. To resolve this, CKKS employs a fixed-point arithmetic approach by introducing a large scaling factor $\Delta$. The encoding process transforms a user-provided message vector $z$ into a plaintext polynomial $m(X)$ by first applying the inverse canonical embedding ($\sigma^{-1}$) to the vector to obtain polynomial coefficients, scaling these coefficients by $\Delta$, and finally discretizing them to the nearest integers. Formally, the encoding function is expressed as:
\begin{equation}
    m(X) = \sigma^{-1}\left( \lfloor \Delta \cdot \pi^{-1}(z) \rceil \right) \in \mathcal{R}
\end{equation}
where $\pi^{-1}$ maps the message vector to the full canonical embedding space, and $\lfloor \cdot \rceil$ denotes the rounding operation to the nearest integer coefficients. This scaling procedure effectively lifts the fractional values into the integer domain, ensuring that significant figures are preserved far above the rounding noise. It is important to note that this rounding step introduces the initial approximation error, which is treated as part of the numerical error inherent to the scheme.

The cryptographic security of the CKKS scheme is founded on the hardness of the RLWE problem, which ensures that recovering the secret key from noisy linear equations over polynomial rings is computationally infeasible. The key generation process begins with sampling a secret key $s(X)$ from a distribution of sparse polynomials with coefficients in $\{0, \pm 1\}$ and a fixed Hamming weight $h$ (e.g., $h=64$). The motivation for employing a sparse secret key, rather than a uniform one, is primarily computational efficiency; sparsity significantly reduces the complexity of polynomial multiplications during homomorphic evaluation without compromising security under appropriate parameter selection. The public key is then constructed to mask this secret. A polynomial $a(X)$ is sampled uniformly at random from the ring arithmetic modulo $q_L$, and a small error polynomial $e(X)$ is sampled from a discrete Gaussian distribution. The public key is computed as the pair $pk = (b, a) = (-a \cdot s + e, a) \pmod{q_L}$. Here, the inclusion of the small error term $e$ is the critical security mechanism; without it, the linear relationship between $b$ and $a$ would allow an adversary to trivially solve for $s$ using Gaussian elimination. The error $e$ makes the pair $(b, a)$ computationally indistinguishable from a uniform random pair.

To encrypt a plaintext polynomial $m(X)$, the scheme generates a ciphertext that computationally hides the message while preserving its algebraic structure for homomorphic operations. The encryption algorithm first samples a random key polynomial $v$ and two small error polynomials $e_0, e_1$. The ciphertext vector $\mathsf{ct} = (c_0, c_1)$ is computed via the linear relation:
\begin{equation}
    \mathsf{ct} = v \cdot pk + (m + e_0, e_1) = (v \cdot b + m + e_0, \ v \cdot a + e_1) \pmod{q_L}
\end{equation}
This construction provides semantic security (IND-CPA) because the random polynomial $v$ ensures that encrypting the same message multiple times yields different, indistinguishable ciphertexts. The decryption structure is derived by computing $c_0 + c_1 \cdot s$, which eliminates the masking terms and yields $m + (v \cdot e + e_0 + e_1 \cdot s)$. In the specific context of CKKS, the motivation for this structure differs from that of exact HE schemes. Rather than considering the residual term $(v \cdot e + e_0 + e_1 \cdot s)$ as a corruption that must be completely removed, CKKS treats it as an intrinsic numerical error. Since the message $m$ has been scaled by a large factor $\Delta$ during encoding, this encryption noise remains bounded within the least significant bits of the plaintext, effectively acting as a negligible approximation error similar to floating-point round-off.

Once data is encrypted into ciphertexts, the CKKS scheme allows for arithmetic operations to be performed directly on the encrypted vectors. For homomorphic addition, the process is straightforward element-wise addition of the ciphertext polynomials. Given two ciphertexts $\mathsf{ct}_1 = (c_{0,1}, c_{1,1})$ and $\mathsf{ct}_2 = (c_{0,2}, c_{1,2})$, the sum is computed as $\mathsf{ct}_{add} = (c_{0,1} + c_{0,2}, c_{1,1} + c_{1,2}) \pmod{q_L}$. The motivation for this simplicity lies in the linear structure of the decryption function; adding the ciphertexts results in the addition of the underlying messages and the summation of their respective error terms. Since the error grows linearly (i.e., $\varepsilon_{new} \approx \varepsilon_1 + \varepsilon_2$) while the message is scaled by a large factor $\Delta$, the relative impact of the additive noise remains negligible, preserving the precision of the result without further adjustment.

Homomorphic multiplication, however, presents significant challenges regarding both ciphertext dimension and message scale. Mathematically, multiplying two ciphertexts corresponds to the tensor product of their decryption equations, $(c_{0} + c_{1}s)(d_{0} + d_{1}s)$, which produces a quadratic polynomial in terms of the secret key: $k_0 + k_1 s + k_2 s^2$. This expansion introduces the first issue: the ciphertext size grows from two polynomials to three. To address this, CKKS employs a process called \textit{Relinearization}. The evaluator uses a public evaluation key ($evk$), which is essentially an encryption of $s^2$, to homomorphically transform the quadratic term $k_2 s^2$ back into a linear form. The motivation for relinearization is to compress the ciphertext back to the standard size of two elements, ensuring that the output size remains constant regardless of the circuit depth and is compatible with the original secret key for future operations.

The second and more critical issue in multiplication is the scale explosion. Since each input message $m$ is encoded as $\Delta \cdot m$, their product results in a message scaled by $\Delta^2$. Furthermore, the encryption noise is also multiplied by the message magnitude, causing the error to grow significantly. To manage this, CKKS introduces its signature operation: \textit{Rescaling}. This procedure divides the ciphertext by an integer $p$ (typically chosen such that $p \approx \Delta$) and switches the modulus from the current level $q_l$ to the next level $q_{l-1}$. Formally, the rescaled ciphertext is computed as:
\begin{equation}
    \mathsf{ct}' = \left\lfloor \frac{q_{l-1}}{q_l} \cdot \mathsf{ct}_{mult} \right\rceil \pmod{q_{l-1}}
\end{equation}
The motivation for rescaling is twofold: it serves as the cryptographic analog of normalizing the significance in floating-point arithmetic. First, it reduces the message scale from $\Delta^2$ back to $\Delta$, preventing the values from overflowing the modulus bound. Second, and crucially, it reduces the accumulated noise by a factor of $\Delta$. By cutting down the noise proportionally to the message, rescaling preserves the signal-to-noise ratio, ensuring that the relative precision of the computation is maintained throughout deep circuits. This modulus consumption strategy effectively trades the available multiplicative depth (the chain of moduli) for numerical stability.

We further examine how the choice of CKKS scale (in bits) affects the fidelity and performance of our TIP on the MNIST task, as shown in \Cref{tab:ablation_scale}. The CKKS scale determines the fixed‐point precision with which real‐valued gradients and Hessian‐vector products are represented under encryption: too small a scale leads to severe quantization error, while an excessively large scale can incur unnecessary overhead without significant accuracy gains.

\begin{table}[ht]
\centering
\resizebox{0.9\textwidth}{!}{%
\begin{tabular}{ccccc}
\toprule
\textbf{Scale} & \textbf{Pearson Correlation} & \textbf{Mean Absolute Error}  & \textbf{Running Time Per Sample} \\ \midrule 
30             & 0.3338                       & 2.286379e-02                                      & 0.1474                         \\ 
31             & 0.5807                       & 2.502078e-03                                      & 0.1480                         \\ 
32             & 0.8227                       & 5.438352e-04                                      & 0.1466                           \\ 
33             & 0.9458                       & 2.733356e-04                                      & 0.1473                           \\ 
34             & 0.9854                       & 1.921560e-04                                      & 0.1472                           \\ 
35             & 0.9963                       & 1.839612e-04                                      & 0.1470                           \\ 
36             & 0.9991                       & 5.085630e-04                                      & 0.1462                           \\ 
37             & 0.9998                       & 1.305089e-04                                      & 0.1475                           \\ 
38             & 0.9999                       & 5.862073e-05                                      & 0.1463                           \\
39             & 1.0000                       & 9.867892e-06                                      & 0.1478                           \\
40             & 1.0000                       & 1.252196e-05                                      & 0.1468                           \\ \bottomrule
\end{tabular}
}
\caption{Ablation Study on Different Scale}
\label{tab:ablation_scale}
\end{table}

On the lowest scale we tested, the Pearson correlation between the FHE and the plaintext influence scores is only $0.33$, and the mean absolute error is on the order of $2.5\times10^{-3}$. Further increments in scale yield steadily diminishing quantization error: by 34 bits we already achieve a correlation of 0.9854 and MAE of $1.9\times10^{-4}$, and by 38–39 bits we effectively recover perfect correlation with MAE below $10^{-5}$.
Importantly, this dramatic improvement in numerical fidelity comes at almost no additional cost in runtime. In other words, once the encryption parameters are chosen to ensure adequate precision (roughly 36 bits or higher in our experiments), the overhead of packing and linear‐homomorphic operations dominates.

\subsubsection{Example of Secure Computing under Homomorphic Encryption}
\label{app:exampleckks}

Below are illustrative examples of how to use CKKS to perform addition and multiplication operations. Consider a scenario where Alice holds a private input $m_1$ and Bob holds a private input $m_2$. They wish to compute the sum ($m_1 + m_2$) and the product ($m_1 \cdot m_2$) without revealing their respective values to each other. Traditional HE schemes were originally designed for exact integer arithmetic, which poses challenges for real-world data analysis involving floating-point numbers. The intuition is that while generating a polynomial equation from secret parameters is easy, solving for those secrets becomes computationally infeasible when a small, random noise is added to mask the linear relationship.

Under the CKKS setting, first, Alice generates cryptographic keys using a polynomial ring. She samples a sparse polynomial as secret key $s(X)$. She then generates a public key $pk$, which mathematically looks like $-a \cdot s + e$. Here, $a$ is a uniform random polynomial, and $e$ is the small RLWE noise. Because $e$ masks the relationship between $a$ and $s$, the public key reveals no information about $s$. Alice also generates an evaluation key ($evk$) to assist in multiplication.

Since the encryption scheme works on integers, the real numbers $m_1$ and $m_2$ must be converted. A large scaling factor $\Delta$ is used to map them into integer polynomials $\mu_1$ and $\mu_2$:
\begin{equation}
    \mu(X) \approx \Delta \cdot m
\end{equation}
This step lifts the significant figures of the data far above the noise floor, ensuring precision is preserved.

Alice and Bob encrypt their encoded messages $\mu_1$ and $\mu_2$ using the public key. The resulting ciphertexts, $\mathsf{ct}_1$ and $\mathsf{ct}_2$, are vectors of polynomials $(c_0, c_1)$ structured such that:
\begin{equation}
    c_0 + c_1 \cdot s \approx \mu \quad (\text{plus a small encryption noise})
\end{equation}
Even though $\mathsf{ct}_1$ and $\mathsf{ct}_2$ contain the messages, they appear computationally indistinguishable from random noise without the secret key $s$.

With the above settings and pre-processing, to compute the sum, the operation is a straightforward element-wise addition of the ciphertext vectors:
\begin{equation}
    \mathsf{ct}_{\text{add}} = \mathsf{ct}_1 + \mathsf{ct}_2
\end{equation}
When decrypted, this yields $(\mu_1 + \mu_2)$, which corresponds to $\Delta(m_1 + m_2)$. The noise grows linearly but remains negligible relative to the large scaling factor $\Delta$.

To compute the product, the ciphertexts are multiplied using tensor product operations. This results in a raw ciphertext with three components corresponding to $1, s, \text{and } s^2$, and the underlying message scales to $\Delta^2 (m_1 \cdot m_2)$. This introduces two issues: dimension expansion and scale explosion.
\begin{itemize}
    \item \textbf{Relinearization:} The evaluation key ($evk$) is used to compress the three components back into the standard two-component format $(c'_0, c'_1)$, effectively handling the $s^2$ term without revealing $s$.
    \item \textbf{Rescaling:} The ciphertext is divided by $\Delta$. This reduces the message scale from $\Delta^2$ back to $\Delta$ and, crucially, reduces the accumulated noise proportionally.
\end{itemize}

After computation, Bob sends the resulting ciphertext $\mathsf{ct}_{\text{res}} = (c_0, c_1)$ back to Alice. Alice uses her secret key $s$ to compute $c_0 + c_1 \cdot s$. This operation removes the masking polynomials and recovers the noisy encoded message $\mu' \approx \Delta \cdot m_{\text{res}}$. Finally, Alice decodes $\mu'$ and divides by $\Delta$ to recover the precise floating-point result, $m_1 + m_2$, filtering out the negligible encryption noise.

To concretize the cryptographic mechanisms, consider a multiparty scenario involving Alice, Bob, and a Broker: Alice holds a private value $x = 1.2345$; Bob holds a private value $y = 2.5000$; The Broker is tasked to compute the function $f(x, y) = x + y + x \cdot y$ without seeing $x$ or $y$. We set the scaling factor $\Delta = 10^4$. The exact ground truth is $1.2345 + 2.5 + 3.08625 = 6.82075$. The encryption calculation proceeds as follows:

Alice first generates a key pair: a Secret Key ($sk$) effectively used to remove noise during decryption, and a Public Key ($pk$) to inject noise during encryption. Crucially, she also generates a Relinearization/Evaluation Key ($evk$) for scaling and rescaling.
She keeps $sk$ private and sends ($pk$) to Bob, and ($evk$) to the Broker. The $evk$ is essential for the Broker to manage ciphertext size expansion after multiplication without seeing the data.

Then Alice and Bob map their real values to integers by multiplying by $\Delta = 10^4$:
\[
\mu_x = \lfloor 1.2345 \times 10^4 \rceil = 12,345, \quad \mu_y = \lfloor 2.5000 \times 10^4 \rceil = 25,000.
\]
They verify that these integers fit within the message modulus, encrypt them into $Enc_{pk} (x)$ and $Enc_{pk} (y)$ using $pk$, and send them to the Broker.

The Broker first computes the product term. It performs $Enc_{mult} = Enc_{pk} (x) \otimes Enc_{pk} (y)$ using $evk$.
In the underlying plaintext space, the integers are multiplied:
\[
Enc_{mult} = Enc_{pk}(12,345) \otimes Enc_{pk}(25,000) = Enc_{pk}(308,625,000).
\]
The result $308,625,000$ represents the value $3.08625$ scaled by $\Delta^2 = 10^8$. It cannot be directly added to $Enc_{pk} (x)$ or $Enc_{pk} (y)$ because they are at scale $\Delta = 10^4$. That is, the encryption noise $\varepsilon$ is multiplied by the message magnitude, causing it to grow significantly. 
To align the scales, the Broker applies $Enc_{mult}' \leftarrow \mathsf{Rescale}(Enc_{mult})$. This operation divides the underlying message by $\Delta$:
\[
Enc_{mult}' = \lfloor Enc_{pk}(308,625,000) / 10^4 \rceil = Enc_{pk}(30,862) .
\]
This operation restores the scale to $\Delta$ and, critically, reduces the noise by a factor of $\Delta$, keeping the computation stable. Note that the last digit is truncated ($...25 \to ...2$), representing the precision loss. Now that all terms are at scale $\Delta$, the Broker computes the final sum:
\[
Enc_{final} = Enc_{pk} (x) + Enc_{pk} (y) + Enc_{mult}' = Enc_{pk}(68,207).
\]

Unlike multiplication, the encryption noise in addition grows linearly ($\varepsilon_1 + \varepsilon_2$). Since $\Delta$ ($10^4$) provides a large buffer, this additive noise remains in the fractional range (e.g., $< 0.5$) and does not corrupt the integer $Enc_{final}$. Thus, no rescaling is needed after addition.

Finally, the Broker sends $Enc_{final}$ to Alice. She decrypts it using $sk$:
\[
\text{Result} = Dec_{sk} (Enc_{pk}(68,207)) = 68,207,
\]
and decodes by dividing by the scaling factor $\Delta$:
\[
\text{Result}' = 68,207 / 10^4 = 6.8207.
\]
Comparing this to the ground truth ($6.82075$), the absolute error is $0.00005$. This demonstrates that CKKS successfully computes complex mixed operations while maintaining high precision. By choosing a more appropriate scaling factor, CKKS can preserve higher precision in more digits.

\subsection{Dataset-Level Evaluation}
\label{app:additive_theorem}

We detail the theoretical basis for valuing a dataset via the summation of individual influence scores. We follow the framework established by ZAMinfluence \citep{broderick2020automatic} and analyzed by MISS \citep{hu2024most}, which justifies the use of greedy heuristics for subset selection.

Consider a situation of $k$ candidate data points from $S$ being (randomly) selected to form a candidate selling dataset. The classic group IF algorithms assume additivity, which approximates the group effect as the sum of individual influences:
$$
A_{S_k} \approx \sum_{s \in S} \mathcal{I}(z_s, z_{eval}).
$$
where $A$ stands for the total influence score for the candidate selling dataset.
The above assumes that the interaction between data points is negligible. However, \citet{hu2024most} shows that this assumption fails when the interaction term, amplification and cancellation, causes IF to fail. Amplification occurs when points in a subset are highly correlated. And cancellation happens when one point neutralizes the effect of removing another. Thus, a trustworthy way to calculate the Influence-function-based score for a dataset is by heuristic approaches.
For example, \citet{hu2024most} proposed an Adaptive Greedy method, which accounts for interactions between data points. 
Specifically, instead of selecting $k$ points at once based on static scores, the algorithm selects the single most influential sample, effectively removes it from the training set, and refits the model before evaluating the remaining samples. By recalculating scores at each step, the method measures the marginal contribution of each data point relative to the subset already selected, rather than relying on a one-time estimate derived from the full dataset. This dynamic update allows the algorithm to detect and correct for non-additive interactions like amplification and cancellation, where the influence of a point significantly changes after a correlated point is removed.

\subsection{Scalable Influence Computation Details}
\label{app:scalable}

This appendix follows \citet{choe2024your} and details a memory- and compute-efficient gradient projection pipeline (LoGra) for scalable influence computation in large neural networks. Our goal is to compute influence scores of the form:
$$
\mathcal{I}(z_s, z_{\mathrm{eval}}) \;=\; -\, \nabla_\theta \ell\bigl(z_{\mathrm{eval}};\hat{\theta}\bigr)^\top H_{\hat{\theta}}^{-1} \nabla_\theta \ell(z_s;\hat{\theta}).
$$
without materializing per-sample full-parameter gradients or any Hessian matrix directly. The key idea is to (i) project per-sample gradients onto low-dimensional coordinates using the inherent Kronecker structure of backpropagation, and (ii) approximate the curvature inverse in the same projected coordinates via layer-wise Kronecker-factored approximations.

\subsubsection{Theoretical Justification}
\label{app:LoGrajustificaion}

To solve this problem, instead of the full gradient influence, methods like TRAK \citep{park2023trak} project gradients onto a lower-dimensional space using a random projection matrix $P \in \mathbb{R}^{k \times p}$:
$$
\mathcal{I}(z_s, z_{\mathrm{eval}}) \approx (P g_{\mathrm{eval}})^\top (P H P^\top)^{-1} (P g_s).
$$

It reduces the vector size from $p$ (billions) to $k$ (thousands), converting the influence calculation into a manageable vector similarity search. First, we must verify that such dimensionality reduction does not invalidate the influence metric. 
When calculating the influence function, the standard formula is $ \mathcal{I}= g_{\mathrm{eval}}^{\top} H^{-1} g_s$. In practice, a damping term $\lambda I$ is usually added to ensure that the Hessian matrix $H$ is invertible. 
\citet{choe2024your} provide a theoretical justification by demonstrating that the damping term $\lambda I$ used in standard influence functions acts as a soft spectral filter, mathematically bounding the contribution of gradients in directions of low curvature. 

Considering that language modeling falls under the maximum likelihood estimation (MLE) framework, we first replace the Hessian using the Fisher Information Matrix (FIM), and further approximate the FIM with the empirical FIM (EFIM). For a model with parameters $\theta$, the FIM and EFIM are defined as:

\begin{align}
    H
    &=
    \mathbb{E}_{p_\theta(y\mid x)}
    \!\left[
    \nabla \log p_\theta(y\mid x)\,
    \nabla \log p_\theta(y\mid x)^{\top}
    \right] \\
\label{ass:FIM}
    &\approx
    \frac{1}{N}
    \sum_{(x_n,y_n)\in D_{\mathrm{tr}}}
    \left[
    \nabla \log p_\theta(y_n\mid x_n)\,
    \nabla \log p_\theta(y_n\mid x_n)^{\top}
    \right].
\end{align}

Assume $g_{\mathrm{eval}}$ and $g_s$ approximately follow the same distribution. 
Let $\{e_i\}_{i=1}^n$ and  $\{\lambda_i\}_{i=1}^n$  be the eigenvectors and eigenvalues of the Hessian. Denote $Q = [e_1, \cdots, e_n]$ and $\Lambda = \text{diag}(\lambda_1, \cdots, \lambda_n)$. 
Under the assumption in Eq.~\eqref{ass:FIM}, the Hessian is approximated by the empirical FIM, symmetric and positive semi-definite. By the property in MLE:
$$
H \approx \frac{1}{N} \sum \nabla \log p(y|x) \nabla \log p(y|x)^\top = \mathbb{E}[gg^\top].
$$

It also admits an eigendecomposition:
$$
H = Q \Lambda Q^\top = \sum_{i=1}^n \lambda_i e_i e_i^\top.
$$
where $\{\lambda_i\}_{i=1}^n$ are non-negative eigenvalues and $\{e_i\}_{i=1}^n$ are the corresponding orthonormal eigenvectors. The eigenvectors $\{e_i\}_{i=1}^n$ form an orthonormal basis of the parameter space. Since any gradient vector $g \in \mathbb{R}^n$ can be expressed as a linear combination of the basis $\{e_i\}_{i=1}^n$, we represent $g$ as:
$$
g = \sum_{i=1}^n \alpha_i e_i = \sum_{i=1}^n c_i (\sqrt{\lambda_i} e_i).
$$
where $c_i := \frac{\alpha_i}{\sqrt{\lambda_i}} = \frac{e_i^\top g}{\sqrt{\lambda_i}}$. Thus we can derive:

\begin{align*}
    \mathcal{I} &= g_{\mathrm{eval}}^{\top} (H + \lambda I)^{-1} g_s \\
    &= g_{\mathrm{eval}}^\top (Q \Lambda Q^\top + \lambda I)^{-1} g_s \\
    &= g_{\mathrm{eval}}^\top \big( Q(\Lambda + \lambda I)Q^\top \big)^{-1} g_s \\
    &= g_{\mathrm{eval}}^\top Q (\Lambda + \lambda I)^{-1} Q^\top g_s \\
    &= 
    \Big( \sum_i c_{{te},i} \sqrt{\lambda_i} e_i \Big)^\top
    Q(\Lambda + \lambda I)^{-1}Q^\top
    \Big( \sum_i c_{{tr},i} \sqrt{\lambda_i} e_i \Big) \\
    &=
    \begin{bmatrix}
    c_{{te},1}\sqrt{\lambda_1} \\
    \vdots \\
    c_{{te},n}\sqrt{\lambda_n}
    \end{bmatrix}^\top
    (\Lambda + \lambda I)^{-1}
    \begin{bmatrix}
    c_{{tr},1}\sqrt{\lambda_1} \\
    \vdots \\
    c_{{tr},n}\sqrt{\lambda_n}
    \end{bmatrix} \\
    &=
    \sum_{i=1}^n 
    \frac{\lambda_i}{\lambda_i+\lambda}
    \,c_{{tr},i}\,c_{{te},i}.
\end{align*}

This result formally demonstrates that the damping term $\lambda$ acts as a spectral filter. Components where $\lambda_i \gg \lambda$ are preserved (weight $\approx 1$), while components where $\lambda_i \ll \lambda$ are penalized (weight $\approx 0$). This provides the mathematical motivation for LoGra, as it justifies focusing computational resources exclusively on the $k$ largest spectral components through low-rank projection.

Moreover, the variance of the gradient projected onto the $i$-th eigenvector is given by $\mathbb{E}[(e_i^\top g)^2] = e_i^\top \mathbb{E}[gg^\top] e_i \approx \lambda_i$.
Since the projection $e_i^\top g$ equals $c_i \sqrt{\lambda_i}$, it follows that $\mathbb{E}[(c_i \sqrt{\lambda_i})^2] \approx \lambda_i$, which implies $\mathbb{E}[c_i^2] \approx 1$, meaning we can be confident that directions with small eigenvalues contribute almost nothing to the final influence score because the weight $\frac{\lambda_i}{\lambda_i + \lambda}$ will be near zero. Thus, we can safely discard those small components to save memory and compute LoGra without losing the signal of the data valuation.

\subsubsection{Low-rank Gradient Projection}

Given the mathematical motivation of gradient projection approaches to influence functions, we are ready to proceed to project gradients onto a lower-dimensional space. However, in standard TRAK, the projection matrix $P$ itself is still huge ($k \times n$). Even if $k$ is small, multiplying a $k \times n$ matrix by an $n$-sized gradient is still expensive ($O(nk)$), and storing $P$ takes massive memory. To achieve this alignment efficiently without materializing the full matrix $P$, LoGra exploits the structure of backpropagation. The arguments below are at the granularity of each layer, ignoring the cross-layer interactions (block-diagonal assumption) to achieve scalability.

Consider a typical layer in neural networks whose core operation is matrix multiplication. Let $W\in\mathbb{R}^{n_o\times n_i}$ be the weight matrix for the layer. In language modeling, activations carry a sequence dimension $T$; denote the input and output by:
$x_{in} \in \mathbb{R}^{n_i\times T}, x_{out} \in \mathbb{R}^{n_o\times T}.$ The forward pass is:

\begin{equation}
\label{eq:forward}
x_{out} = W x_{in}. 
\end{equation}
 
Let $\mathcal{D} x_{out} \in \mathbb{R}^{n_o\times T}$ be the derivative of the loss with respect to the module output. The backward pass yields:

\begin{equation}
\label{eq:backward}
\text{vec}(\mathcal{D} W) = \sum_{t=1}^{T} x_{in,t} \; \otimes\; \mathcal{D} x_{out,t}, \qquad \mathcal{D} x_{in} = W^\top \mathcal{D} x_{out},
\end{equation}

where $x_{in,t}$ and $\mathcal{D} x_{out,t}$ are the $t$-th token/position columns, $\otimes$ is the Kronecker product, and $\mathrm{vec}(\cdot)$ is vectorization. Equation~\eqref{eq:backward} shows that the per-sample parameter gradient is a sum of Kronecker products between forward activations and backward signals. Notice that the gradient $\mathrm{vec}(\mathcal{D} W)$ obtained during backpropagation is structured as a sum of Kronecker products between forward and backward activations. Thus, instead of using a dense random matrix $P$ (like TRAK), LoGra defines $P$ as a Kronecker product of two smaller matrices: $P = P_i \otimes P_o$, where $P_i \in \mathbb{R}^{k_{in} \times n_{in}}$
projects the input dimension from $n_i$ to $ k_i$, and $P_o \in \mathbb{R}^{k_{out} \times n_{out}}$ projects the output dimension from $n_o$ to $k_o$. 
Under the setting $n_i \approx n_o \approx \sqrt{n}$ and $k_i \approx k_o \approx \sqrt{k}$, where $k\ll n$, applying this structured $P$ to the gradient: 

$$
P \cdot \text{vec}(\mathcal{D}W) = (P_i \otimes P_o) \sum (x_{i,t} \otimes \mathcal{D}x_{o,t}),
$$

Using the identity $(A \otimes B)(C \otimes D) = (AC) \otimes (BD)$:
$$
P \cdot \text{vec}(\mathcal{D}W) = \sum_{t=1}^{T} (P_i x_{i,t}) \otimes (P_o \mathcal{D}x_{o,t}).
$$

From above, by compressing the input vector $x_{in} \in \mathbb{R}^{n_{in}}$ using $P_i$ and the error vector $\mathcal{D}x_{out}  \in \mathbb{R}^{n_{out}}$ using $P_o$, LoGra reduces significantly computational complexity from $O(nk)$ (TRAK) to roughly $O(\sqrt{nk})$. This formulation allows us to populate $P_i$ and $P_o$ with the principal components of the KFAC approximation, thereby satisfying the spectral alignment requirement while maintaining computational tractability.

\subsubsection{Hessian Approximation by KFAC}

From above, the goal of LoGra is to project the massive gradient space into a tiny subspace ($k$). However, if we project randomly, we might lose the most informative directions. 
We used KFAC (Kronecker-Factored Approximate Curvature) for initialization of projection matrices ($P$) to ensure the low-rank projection preserves the most important gradient information. By aligning the projection matrices with the principal components (eigenvectors) of the KFAC approximation, LoGra ensures that the projected gradients retain the most useful signal for calculating influence.

To derive this formally, we begin by acknowledging that computing the exact Hessian $H$ for LLMs is computationally intractable. For a specific layer with weight matrix $W \in \mathbb{R}^{n_{out} \times n_{in}}$, input activation $x_{in} \in \mathbb{R}^{n_{in}}$, and output error gradient $\delta_{out} = \nabla_{x_{out}} \ell \in \mathbb{R}^{n_{out}}$, the gradient with respect to the weights is given by the outer product:
\begin{equation}
    \nabla_{W} \ell = \delta_{out} x_{in}^\top
\end{equation}

In the vectorized form required for the FIM, this becomes the Kronecker product of the activation and the error:

\begin{equation}
    \text{vec}(\nabla_{W} \ell) = x_{in} \otimes \delta_{out}
\end{equation}

Substituting this into the definition of the FIM, the curvature block for this layer is:
\begin{equation}
    H_{layer} = \mathbb{E} \left[ (x_{in} \otimes \delta_{out}) (x_{in} \otimes \delta_{out})^\top \right]
\end{equation}
Using the mixed-product property of Kronecker products, $(A \otimes B)(C \otimes D) = (AC) \otimes (BD)$, we expand the term:
\begin{equation}
    H_{layer} = \mathbb{E} \left[ (x_{in} x_{in}^\top) \otimes (\delta_{out} \delta_{out}^\top) \right]
\end{equation}

Because KFAC assumes that the activations $x_{in}$ and the errors $\delta_{out}$ are statistically independent, the expectation of the Kronecker product factorizes into the Kronecker product of the expectations:

\begin{equation}
    H_{layer} \approx \mathbb{E} [x_{in} x_{in}^\top] \otimes \mathbb{E} [\delta_{out} \delta_{out}^\top] = C_F \otimes C_B
\end{equation}

where $C_F = \mathbb{E} [x_{in} x_{in}^\top] \in \mathbb{R}^{n_{in} \times n_{in}}$ is the forward covariance matrix; and $C_B = \mathbb{E} [\delta_{out} \delta_{out}^\top] \in \mathbb{R}^{n_{out} \times n_{out}}$ is the backward covariance matrix. This decomposition allows us to perform eigendecomposition efficiently. If $C_F$ has eigenvectors $Q_F$ and $C_B$ has eigenvectors $Q_B$, then the eigenvectors of the full approximate Hessian are $Q_F \otimes Q_B$.

Therefore, to capture the directions of maximum curvature, i.e., the principal components of $H_{layer}$, LoGra initializes the input projection matrix $P_i$ with $Q_F^{[1:k_{in}]}$, the top-$k_{in}$ eigenvectors of $C_F$; and the output projection matrix $P_o$ with $Q_B^{[1:k_{out}]}$, the top-$k_{out}$ eigenvectors of $C_B$.
This ensures that the low-rank projection $P = P_i \otimes P_o$ effectively spans the subspace where the influence signal is strongest, minimizing the information loss caused by dimensionality reduction.

\end{document}